\documentclass[10pt,letterpaper]{article}
\usepackage[top=0.85in,left=2.75in,footskip=0.75in]{geometry}

\usepackage{amsmath,amssymb}

\usepackage{changepage}

\usepackage{textcomp,marvosym}

\usepackage{cite}

\usepackage{bbm}

\usepackage{nameref,hyperref}

\usepackage[right]{lineno}

\usepackage[nopatch=eqnum]{microtype}
\DisableLigatures[f]{encoding = *, family = * }

\usepackage[table]{xcolor}

\usepackage{array}

\newcolumntype{+}{!{\vrule width 2pt}}

\newlength\savedwidth

\raggedright
\usepackage[aboveskip=1pt,labelfont=bf,labelsep=period,justification=raggedright,singlelinecheck=off]{caption}

\makeatletter
\renewcommand{\@biblabel}[1]{\quad#1.}
\makeatother

\usepackage{lastpage,fancyhdr,graphicx}
\usepackage{epstopdf}
\fancyheadoffset[L]{2.25in}
\fancyfootoffset[L]{2.25in}
\usepackage{subcaption}

\usepackage{float}
\usepackage{tabularx}
\usepackage{tikz}
\usetikzlibrary{arrows.meta,positioning,fit,shadows.blur,calc}
\tikzset{
  box/.style={draw, rounded corners, align=center, inner sep=4pt, outer sep=2pt, fill=gray!5, text width=33mm},
  fat/.style={draw, rounded corners, very thick, align=center, inner sep=6pt, outer sep=2pt, fill=gray!10, text width=35mm},
  group/.style={draw, rounded corners, inner sep=6pt, fill=gray!3, text width=46mm}
}
\usepackage{standalone}

\begin{document}
\vspace*{0.2in}

\begin{flushleft}
{\Large
\textbf{Causal interventions in bond multi-dealer-to-client platforms} 
}
\newline
\\
Paloma Mar\'in\textsuperscript{1,2},
Sergio Ardanza-Trevijano\textsuperscript{2,3},
Javier Sabio\textsuperscript{1}
\\
\bigskip
\textbf{1} Corporate \& Investment Banking, BBVA, Madrid, Madrid, Spain
\\
\textbf{2} Institute for Data Science and Artificial Intelligence, University of Navarra, Pamplona, Navarra, Spain
\\
\textbf{3} Department of Physics and Applied Mathematics, University of Navarra, Pamplona, Navarra, Spain
\\
\bigskip
\end{flushleft}


%
\section*{Abstract}
The digitalization of financial markets has shifted trading from voice to electronic channels, with Multi-Dealer-to-Client (MD2C) platforms now enabling clients to request quotes (RfQs) for financial instruments like bonds from multiple dealers simultaneously. In this competitive landscape, dealers cannot see each other's prices, making a rigorous analysis of the negotiation process crucial to ensure their profitability. This article introduces a novel general framework for analyzing the RfQ process using probabilistic graphical models and causal inference. Within this framework, we explore different inferential questions that are relevant for dealers participating in MD2C platforms, such as the computation of optimal prices, estimating potential revenues and the identification of clients that might be interested in trading the dealer's axes. We then move into analyzing two different approaches for model specification: a generative model built on the work of (Fermanian, Guéant, \& Pu, 2017); and discriminative models utilizing machine learning techniques. Our results show that generative models can match the predictive accuracy of leading discriminative algorithms such as LightGBM (ROC-AUC: 0.742 vs. 0.743) while simultaneously enforcing critical business requirements, notably spread monotonicity.

\section{Introduction}
\label{Introduction}

Modern financial markets are increasingly mediated by electronic platforms that automate the negotiation and execution of trades. While highly liquid instruments such as equities are typically traded on order-driven venues using continuous double auctions and limit order books, these mechanisms are less effective for instruments that are not as actively traded, such as corporate and government bonds. Unlike equities, where a company issues a single line of stock, bond issuers may maintain hundreds of distinct securities with varying maturities, coupons, and seniorities. This proliferation of instruments fragments liquidity and reduces the probability of matching counterparties through centralized order books.

To address these limitations, quote-driven protocols—particularly the Request-for-Quote (RfQ) mechanism—are commonly used. In this setting, Multi-Dealer-to-Client (MD2C) platforms have emerged as the dominant architecture for institutional bond trading. These platforms allow clients to simultaneously solicit quotes from multiple dealers and, within a short response window, choose whether to trade at the best available price \cite{JSG2024}. The process is partially observable: the winning dealer is informed of the second-best quote (the “cover price”), while the other dealers receive feedback on their ranking and whether a trade occurred\footnote{In some cases, the other dealers cannot even know if the trade occurred (a {\it missed} RfQ) or not (a {\it passed} RfQ)}. This structure fosters competition but creates a highly strategic environment for pricing, in which dealers must optimize quotes under uncertainty about competitors’ prices and client preferences.

For dealers, RfQ pricing entails balancing the probability of winning a trade with expected profitability and inventory risk. Quoting too aggressively may increase hit probability but reduce margins or expose the dealer to adverse selection and post-trade market movements. Additionally, dealers seek to extract commercial insights from the process, such as identifying clients who may be receptive to trading their axes—pre-existing positions the dealer wishes to buy or sell. However, optimizing these decisions from historical data is challenging due to confounding factors. Pricing policies are often shaped by client segmentation or product-specific rules, meaning that the historical relationship between quotes and outcomes may not reflect the causal effect of intervening on prices or initiating commercial actions.

To address the first research question of this study —namely, assessing the potential impact of confounding factors in the historical datasets used by dealers for decision-making—this paper introduces a unified framework based on probabilistic graphical models \cite{koller2009probabilistic, murphy2013machine, denev2015probabilistic} and causal inference \cite{pearl2016causal, Pearl09}. Graphical models allow us to explicitly encode the dependencies among variables involved in the RfQ workflow, including observed and latent factors. Causal inference techniques then enable the analysis of interventions—such as setting specific quotes or initiating outreach to clients—in a manner that accounts for confounding biases and supports counterfactual reasoning. In particular, we focus on three high-impact dealer use cases: (i) optimal pricing to maximize expected profits or satisfy commercial targets, (ii) revenue potential estimation under fixed pricing policies, and (iii) identifying likely matches between axes and client demand. 

Our work thus continues a recent trend in which causal inference methods are being popularized across economics and finance, although their use in financial markets—traditionally dominated by correlation-based analyses—remains at an early stage. Early applications in empirical economics have focused on identifying causal effects in macroeconomic and policy contexts. For example, Baiardi and Naghi~\cite{baiardi2024value} revisited classical empirical financial economics studies using Double Machine Learning~\cite{chernozhukov_doubleML}, a methodology that integrates causal inference with modern machine learning to estimate structural relationships more reliably. Similarly, \cite{kumar2024unveiling} have examined the causal impact of interest rate changes on fixed income funds through Double ML estimation, illustrating the growing use of causal ML to evaluate policy transmission mechanisms. More recently, causal methods have been applied directly to financial markets: Oliveira et al.~\cite{oliveira2024causality} have proposed causality-inspired forecasting models that remain robust under distributional shifts, demonstrating improved performance in volatile investment environments. Extending this line of research, López de Prado and Zoonekynd~\cite{lopezdeprado2025causality} advocate for a causal investing paradigm grounded in counterfactual reasoning and experimental design, framing causal inference as a foundation for more reliable discovery of investment factors and strategies.

The second primary research question of this study concerns the comparative effectiveness of different modeling paradigms for representing the RfQ negotiation process. Specifically, we aim to assess the relative advantages of generative models that incorporate the internal mechanics of dealer–client interactions versus discriminative models that directly learn predictive mappings from data. To this end, we develop two complementary strategies to instantiate the graphical model.

The generative model, inspired by the work of Fermanian, Guéant, and Pu~\cite{fermanian2016behavior}, explicitly represents the negotiation mechanism underlying RfQ interactions. This approach naturally enforces economically consistent behaviors—such as price monotonicity, whereby clients prefer more favorable quotes—and seamlessly integrates post-trade information like the cover price. In contrast, discriminative models, such as logistic regression and gradient-boosted decision trees~\cite{bishop2007}, directly estimate outcome probabilities conditional on observed RfQ features. These models offer greater flexibility and scalability but may struggle to encode economic constraints or effectively leverage post-trade feedback. Comparing these two approaches enables us to evaluate the trade-off between structural interpretability and predictive performance within the causal inference framework.

To evaluate these approaches, we benchmark their performance on a proprietary dataset of BBVA’s RfQ activity in European Government Bonds (EGBs). We frame the core estimation problem as a binary classification task—predicting trade outcomes conditional on quotes and market context—and assess each model’s ability to support decision-making tasks under a causal perspective. While the analysis focuses on EGBs, the methodology is general and should apply to any asset class traded via the RfQ protocol.

The remainder of the paper is structured as follows. Section II introduces the probabilistic graphical model underlying the RfQ process. Sections III and IV apply causal inference techniques to evaluate key business interventions within M2DC markets, focusing on optimal pricing, revenue estimation, and axe matching. Section V details both generative and discriminative modeling strategies, with emphasis on practical considerations for estimation and inference. Section VI compares these approaches in the context of a specific inference task—the {\it hit probability model}—using predictive performance metrics. Finally, Section VII summarizes the main findings and outlines directions for future research.

\section{A causal graphical model for the RfQ process}
\label{GraphicalModel}

\begin{figure}[H]
\centering
\includegraphics[scale=0.6]{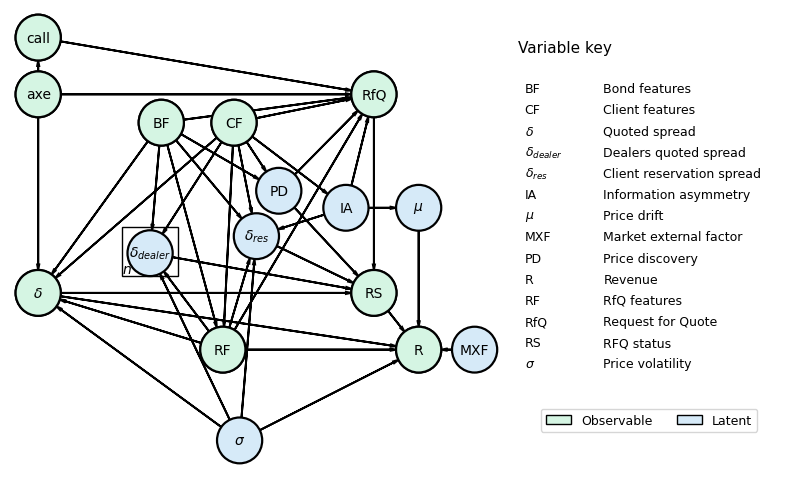}
\caption{\textbf{Causal graphical model for the RfQ process}. Green nodes correspond to observable variables, whereas blue ones are latent variables. A detailed description of each of the variables is provided in the main text \label{fig1}}
\end{figure}

We model the full RfQ process as a graphical model depicted in Fig~\ref{fig1}. This model captures key components of the RfQ process, making it well-suited for pricing analysis as well as other commercial applications, such as client targeting for axe recommendations and estimating a dealer’s expected profitability following a transaction.

The model is composed of the following elements. Some of them are known pre-trade, others post-trade depending on the result of the RfQ, and some are just simply non-observable (latent) variables:
\newline
\newline
\textbf{RfQ}: A Request for Quote sent by a client to a set of dealers via a MD2C platform. This is a binary random variable that specifies the arrival or not of an RfQ. RfQs have features that we model generically with the variable \textbf{RF} in the model, for instance the time $t$ of the RfQ, the side $s$ (buy or sell)\footnote{We consider the side from the perspective of the dealer, i.e. a buy RfQ means that the dealer buys and the client sells, and a sell RfQ means that the dealer sells and the client buys}, the volume $v$ and the number of dealers $n$ in competition, which the client can select when sending the RfQ (actual platforms tend to cap this to a maximum). The initiation of an RfQ may be driven by exogenous client factors—such as the specific trading strategy being pursued by the client—which lie outside the scope of our model. However, in some cases, it may also result from a commercial intervention by the dealer’s sales force, which we represent with the variable \textbf{call}. Additional factors that may influence a client's decision to request a quote, and which we explicitly incorporate into the model, include:
\begin{itemize}
        \item {\bf Price discovery (PD)}: The client wants to find out what is the fair price of the instrument but is not willing to trade. This is a latent variable since the dealer does not know the intention of the client when the RfQ is submitted.
        \item {\bf Client information asymmetry (IA)}: If the client has relevant information about the short-term directionality of the market \cite{ohara1995market, CarteaSanchez2025}. There are different reasons why a client could have information asymmetry. From a causal graph structure, we differentiate between two: 1) the client trading activity represents a potential source of market impact, but the RfQ submitted to the dealer often reveals only a subset of the total order. In this case, the information asymmetry impacts the market, which we model as a price {\bf drift ($\mu$)} variable, and the direct causal effect: $IA \rightarrow \mu$,  2) the client has insider information or advanced analytics models that are relevant for predicting market directionality, again modeled as a price drift variable. In this case, though, there is not a direct link between IA and drift, since the client uses an estimation of the drift, but does not know the actual drift after trading. Since the first case is more general, we will consider it in the rest of the paper, keeping in mind that extra simplifications of the graphical model can be achieved in the second situation. Notice that information asymmetry is closely related to adverse selection effects, a central concern in market microstructure. Recent advances, such as~\cite{barzykin2025optimal}, provide new evidence and models for understanding toxic flow and its implications for dealer profitability.
        \item {\bf Axe}: The dealer makes public an interest in offering a discounted trade (it could be a bid or an ask), which makes the client more inclined to request a quote. 
        \item {\bf Client features (CF)}: The client identity itself, but also more general characteristics that might reflect common patterns of behavior between clients, e.g. industry, geography or ratings. For instance, we expect to find a different behavior from a hedge fund than a central bank, although trading behavior might sometimes be more influenced by the specific trader than the institution. 
        \item {\bf Bond features (BF)}: The bond identifier itself, but also financial characteristics like coupon, yield, maturity, sensitivity to interest rates (for bonds the DV01 or Dollar Value of 01) or variables linked to liquidity conditions like market bid-ask spreads, volatility and the average number of dealers quoting RfQs in this bond.
\end{itemize}
Notice that RfQ features might be dependent on the client requesting the quote or the instrument being priced, and the model incorporates these dependencies with direct causal arrows in the graphical model.
\newline
\newline
\textbf{RfQ status (RS)}: The final status of the RfQ, which can be grouped in three relevant cases. 
        \begin{itemize}
            \item \textbf{Hit}: When the client trades with the dealer. There are two final status that yield this result:
            \begin{itemize}
                \item[] \textbf{Done}: The dealer proposed the best price, with the rest of the dealers quoting worse prices.
                \item[] \textbf{Tied Done}: There was at least other dealer quoting the same price, but the platform mechanism did not assign the trade to these other dealers.
            \end{itemize}
            \item \textbf{Missed}: When the client trades with other dealer. The relevant status grouped in this case are:
            \begin{itemize}
                \item[] \textbf{Tied Traded Away}: When the dealer loses the RfQ but the price quoted matched the one from the best dealer.
                \item[] \textbf{Covered}: When the dealer loses the RfQ but quoted the second best price among all dealers.
                \item[]\textbf{Other Traded Away}: When the dealer loses the RfQ and quoted a price worst than the two best ones. 
            \end{itemize}
            \item \textbf{Passed}: When the client does not trade with any dealer. The reason could be that the prices she received were not good enough or that she was just doing price discovery.
        \end{itemize}
   The drivers of the RfQ status are, in our model:
    \begin{itemize}
        \item Naturally, the client's interest in trading, which translates into a \textbf{Request for Quote (RfQ}) being sent to the dealers using a multi-dealer-to-client platform.
        \item If the client is requesting the quote for the sole reason of {\bf price discovery (PD)}, in which case the only result with non-zero probability is passed (also known as walked away).
        \item The {\bf half-spread ($\delta$)} quoted by the dealer. Although dealers quote prices to clients, we will make the hypothesis that client decisions on trading are driven by the spread with respect to a reference mid-price $P_{m}$, for example a platform's composite price: $P=P_{m}+s \delta$, where $s = 1$ when the dealer sells (ask) and $s = -1$ when the dealer buys (bid).
        \item The {\bf spreads quoted by the dealers in competition}, $\delta_{dealer}$. As mentioned, the client can select a number of dealers $n$ to submit the RfQ. Naturally, as the number of competing dealers increases, so does the likelihood that one of them will offer a more favorable price. Notice that most times $n$ is known to every dealer quoting a price, but the prices quoted by other dealers are not known pre-trade. Only when a dealer wins the RfQ and trades with the client does the platform publish the second best price quoted, the cover price. The second best dealer knows that it quoted the cover price, since she receives the final status {\it covered}. However, she does not know from the platform the price quoted by the dealer trading with the client\footnote{This information can be known in some cases post-trade with delay via proprietary repositories like Trax, or public ones like TRACE in the USA or MiFID 2's APAs in Europe.}.
        \item The client's {\bf reservation spread}, $\delta_{res}$: the maximum spread with respect to the mid-price that the client is willing to buy or sell, i.e. $\delta_{res}= s(P_{res}-P_{m})$ where $s$ is the side defined as in the half-spread. If every dealer quotes a worse price than $\delta_{res}$ the final status will be {\it passed}. This is a latent variable that needs to be inferred from the historical behavior of clients. The drivers of this variable are naturally the client features $CF$, the bond features $BF$ and the RfQ features $RF$. Here, we also include specifically the information asymmetry of the client $IA$, since we expect that having such information will make the client more willing to trade with any dealer. A similar effect is expected from the market \textbf{volatility $\sigma$}: for example, a client might be more willing to liquidate a position when volatility is higher, to avoid potential price risk. 
    \end{itemize}
\textbf{Revenue (R)}: The profitability of the trade. There are different measures of revenue used by dealers:
\begin{itemize}    
\item {\it Instantaneous flow value} ($R_0$): A simple revenue per-trade metric is the instantaneous flow-value, which measures the profitability directly as the one captured by the spread $R_0 = v \delta$ at transaction time. The problem of this measure is that it does not take into account the profits or losses incurred until an opposite trade is closed. 
\item {\it Round-trip revenue} ($R_{rt}$): An issue with instantaneous flow value is that it does not take into account the profit or losses incurred until an opposite trade is done. Such measure requires to identify a round-trip of trading (buy and sell, or vice-versa) and compute $R_{rt} = s v (P_{t_{rt}} - P_t) = v \delta_{t_{rt}} - v \delta_{t} + s v (P_{m,t_{rt}} - P_{m,t})$, where $t$ indicates the time when the RfQ was closed. If we use a simple model for price dynamics, namely a Brownian motion, then $P_{m,t_{rt}} - P_{m,t} = \mu(t_{rt}-t) + \sigma (W_{t_{rt}} - W_t)$, where $\mu$ is the {\bf drift}, $\sigma$ the {\bf volatility}, $t_{rt}$ is the round-trip time, and $W_t$ is a Wiener process, so that its time difference follows a normal distribution $W_{t_{rt}} - W_{t} \sim N(0, t_{rt} - t)$. The integrated Wiener process accounts in our model for the influence of {\bf market external factors (MXF)} in revenues. As discussed above, we also introduced a potential causal dependency between the client information asymmetry (IA) and the drift, to capture situations in which the client trading's activity produces a market impact. A point to note is that $\mu$ and $\sigma$ represent the actual post-trade realized drift and volatility, which are not known a priori by either the client or the dealers. These are latent variables that can only be inferred from post-trade data. In our model, however, we treat volatility as a relatively predictable quantity \cite{andersen2006volatility}; therefore, we assume that pre-trade and post-trade volatility are effectively the same variable. This volatility can potentially influence both the dealers' pricing policies, $\delta$ and $\delta_{\text{dealer}}$, and the client's reservation price, $\delta_{\text{res}}$.
Drift, on the other hand, is widely recognized as much harder to forecast, and we regard any attempt at prediction as unreliable unless the client possesses asymmetric information, as previously discussed. Consequently, the post-trade drift does not affect the dealers' pricing decisions.
\item {\it End of day flow value} ($R_T$): Since round-trips of buying and selling might involve large time scales for illiquid bonds, dealers also tend to compute revenues at a given time-horizon, e.g. end of day, by using market mid-prices to value inventories. In this case, hence, the revenue is computed as $R_T = v \delta_{t} + s v (P_{m,T} - P_{m,t})$, where $T$ is the end of day time. Liquidity penalties can sometimes be included in this formulation to penalize unrealized mark-to-market profits, reflecting the potential cost of carrying illiquid assets.
\item {\it Short-term flow value ($R_{t+h}$)}: Another common revenue measure is using a fixed time window $h$ after trading, which will depend on the liquidity of the bond but is typically on the scale of seconds or minutes. Revenues are then computed as well with respect to the mid-price at this time, namely  $R_{t+h} = v \delta_{t} + s v (P_{m,t +h} - P_{m,t})$. This approach can be helpful to identify clients trading with information asymmetry, since the impact of such information in the profitability of the dealer will be easier to identify in a short-term horizon, before it gets overridden by market external factors (MXF).
\end{itemize}
Notice that we have not taken into account potential hedging activity of the dealer when measuring revenues. The model can be extended to cover those cases, by incorporating the cost and market dynamics of hedging instruments. 

A relevant non-trivial dependence in our graphical model refers to the determinants of pricing policies, both the dealer's one, $\delta$, and those of the dealers in competition $\delta_\text{dealer}$. Since the client identity is known by the dealers quoting in MD2C platforms, their pricing policies tend to incorporate specific features of clients, as well as drivers of the demand and price sensitivity of clients, for example bond features like liquidity conditions or relative value, and RfQ features like the number of dealers in competition. In the model, such dependence is reflected in causal arrows from $BF$ and $CF$, as well to $RF$, to $\delta$ and $\delta_\text{dealer}$. Finally, if the dealer has an axe in the bond, pricing will be also accommodated to skew prices in favor of trading, hence the arrow $\text{axe} \rightarrow \delta $. As we will see, these causal relationships are critical for accurately predicting the effects of interventions, particularly in pricing. Ignoring them can bias optimal pricing models, as these variables introduce spurious dependencies between the quoted price and the RfQ outcome. In the language of Causal Inference, they are known as confounders \cite{pearl2016causal, Pearl09}.

\section{Causal interventions and predictions in the graphical model}
\label{CausalInterventions}

Arguably, the primary application of models in a business context is to support and optimize decision-making. This generally results in specific business actions (or interventions, using the Causal Inference terminology), namely setting prices, launching marketing campaigns, recommending products, and so on \cite{katsov_algo_marketing}. As established over the past few decades by the field of Causal Inference \cite{pearl2016causal, Pearl09}, traditional probability theory is not well suited to address interventional questions. This is because it relies on estimating historical correlations, which may not hold when the objective is to assess the potential impact of an action or intervention on business metrics. To address such questions, we must turn to the formalism of do-calculus \cite{pearl2016causal, Pearl09} to determine whether the effect of an intervention can be identified from historical data. If not, it becomes necessary to design experiments—specifically, randomized controlled trials (RCTs), commonly implemented as A/B tests—to empirically measure the impact of the intervention \cite{Kohavi_trustworthy}). In this context, graphical models play a central role in applying causal inference to practical problems, as they can significantly simplify the underlying mathematical analysis. In the following, we examine several relevant interventions on the graphical model of the RfQ process introduced in the previous section.

\subsection*{Optimal pricing}

In the context of MD2C platforms, dealers must develop pricing strategies to respond to client RfQs, aiming to maximize business profitability while potentially achieving key commercial targets, for instance minimum hit-and-miss ratios\footnote{The hit-and-miss ratio measures the share of RfQs executed with a specific dealer out of all RfQs that resulted in a trade, i.e., excluding RfQs where the client opted not to trade.}. The simplest optimal strategies aim to maximize the expected revenue of each trade independently, without accounting for the risk associated with holding inventory until it is offset by a subsequent trade. More sophisticated approaches, however, consider not only the current RfQ but also potential future requests. In what follows, we analyze some of these strategies through the lens of causal interventions on the graphical model:
\newline
\newline
{\bf Instantaneous flow value optimization}: As the name suggests, this strategy aims to set the spread that maximizes expected revenue, defined in this case as the instantaneous flow value. Since revenues are inherently uncertain prior to execution—here, essentially due to the uncertainty surrounding the client's decision to trade—the dealer must rely on available information to predict expected outcomes: RfQ features like side and volume, client features, bond features, etc. As we will discuss in a later section, proper conditioning is essential for successful optimization, due to the underlying causal relationships among the variables. For now, however, we consider a generic setting in which all information used by the dealer for estimating revenues is included in a conditioning set $\mathcal{Z}_t$: 

\begin{equation}
\delta_{\text{opt}}  =  \text{argmax}_\delta \mathbb{E}[R_0|\text{do}(\delta), \text{RfQ}, \mathcal{Z}_t] = \text{argmax}_\delta \mathbb{E}[v \delta 1_{\text{hit}} |\text{do}(\delta), \text{RfQ}, \mathcal{Z}_t]
\end{equation}

Here, $t$ denotes the time of the RfQ, and $\delta$ represents the half-spread quoted relative to the mid-price. In this formulation, we assume that the mid-price is an exogenous variable, mutually agreed upon by both the client and the dealer. As such, it does not influence the outcome of the RfQ and can therefore be factored out of the analysis. We have also used the simplified notation $1_{RS = \text{hit}} \equiv 1_{\text{hit}}$. The $\text{do}$-operator indicates an intervention on the spread—that is, we are not merely concerned with the historical conditional distribution of RfQs won given the spreads, but rather with the interventional distribution. This distinction is crucial, as it requires a causal, rather than purely associative, relationship between the variables. The optimal pricing formula is easily derived. Let us write: 

\begin{equation}
\mathbb{E}[v \delta 1_{\text{hit}}|\text{do}(\delta), \text{RfQ}, \mathcal{Z}_t] = v \delta P(\text{hit} |\text{do}(\delta), \text{RfQ}, \mathcal{Z}_t) = v \delta f(\delta)
\end{equation}

where $P(\text{hit} |\text{do}(\delta), \text{RfQ}, \mathcal{Z}_t) \equiv f(\delta)$ is called the {\it hit probability}, i.e. the probability that the client will trade the RfQ with the dealer, given the price quoted and the context of the RfQ. Hence:

\begin{equation}
\delta_{\text{opt}} = -\frac{f(\delta_\text{opt} )}{f'(\delta_\text{opt} )}
\end{equation}

which is an implicit formula for the optimal spread for a general functional form for $f(\delta)$. 

A central insight from Causal Inference theory~\cite{pearl2016causal} is that unbiased estimation of the hit probability model critically depends on selecting an appropriate conditioning set. In many cases, a minimal subset of variables can be identified that ensures the validity of traditional probabilistic estimation methods while preserving causal interpretability. Importantly, including all available features indiscriminately can introduce spurious correlations, distorting the estimation of causal effects.

This issue is particularly relevant when the hit probability model is used to inform optimal pricing decisions. In such cases, dealers must isolate the causal effect of their intervention—specifically, the setting of the spread—on the outcome of the RfQ. This requires controlling for confounding variables that influence both the spread and the RfQ outcome, but are not part of the causal pathway. The distinction between standard conditional probabilities and those involving the $\text{do}$-operator hinges precisely on this point. As will be discussed in the next section, tools from Causal Inference—particularly the \textit{back-door criterion}~\cite{pearl2016causal}—provide a principled way to identify the minimal set of variables required for valid causal effect estimation.

In the context of the hit probability model, this minimal conditioning set includes the volatility $\sigma$ (which, as previously noted, we treat as observable in our framework), along with the RfQ features ($RF$), bond features ($BF$), and client features ($CF$). When using this conditioning set, we can write:
\begin{equation}
P(\text{hit} |\text{do}(\delta), \text{RfQ}, \sigma, RF, BF, CF) = P(\text{hit} |\delta, \text{RfQ}, \sigma, RF, BF, CF)
\end{equation}
Intuitively, the key issue is that dealers do not quote spreads based solely on external, context-independent drivers. Instead, they actively incorporate information from the market environment, the RfQ itself, the characteristics of the instrument, the identity of the client, and the competitive landscape, among others, when setting prices. This practice introduces correlations between historical spreads and RfQ outcomes that may not reflect causal relationships and, therefore, may be irrelevant when determining optimal spreads aimed at maximizing expected revenues.

While a minimal set of variables is required to estimate the hit probability model accurately, this does not imply that the dealer must adopt {\it price segmentation} strategies. Nonetheless, when feasible, such strategies can yield higher revenues than {\it global pricing strategies}, as extensively discussed in classical microeconomics and marketing literature \cite{varian10, katsov_algo_marketing}. If the conditioning set includes fewer variables than necessary, the back-door criterion can still be applied by integrating over the distribution of the omitted variables.
\newline
\newline
{\bf Utility maximization and transactional risk}: This simple strategy can be enriched by introducing a risk-return trade-off, namely the transactional risk \cite{Gu_ant_2012}, by maximizing the expected utility of the flow value. Transactional risk arises from the trade-off between low margin, frequent transaction strategies and high margin, less frequent transaction ones, the latter having larger margins per trade but more risk of a less uncertain stream of revenues. In the language of interventions, this reads:

\begin{eqnarray}
\delta_\text{opt} = \text{argmax}_\delta \mathbb{E}[U(v \delta 1_{\text{hit}})|\text{do}(\delta),\text{RfQ}, \mathcal{Z}_t] = \\
\text{argmax}_\delta \left(U(v \delta) f(\delta) + U(0) (1-f(\delta)) \right)
\end{eqnarray}

where $U(\cdot)$ is a utility function. The optimal pricing formula is again an implicit one on the spread, namely:

\begin{eqnarray}
    U'(v \delta_{opt}) = -\frac{1}{v} \frac{f'(\delta_{opt})}{f(\delta_{opt})} \left(U(v \delta_{opt}) - U(0)\right)
\end{eqnarray}

Using the popular exponential utility function $U(x) = 1-\exp{(-\gamma x)}$, the result simplifies to:

\begin{eqnarray}
\delta_\text{opt}  = \frac{1}{\gamma v}\log\left(1-\gamma v \frac{f(\delta_\text{opt})}{f'(\delta_\text{opt})}\right)
\end{eqnarray}

where $\gamma$ is called the risk-aversion parameter. We see again how the hit probability model is central to this optimal pricing strategy. The risk-averse dealer ($\gamma > 0$) will sacrifice profits per trade to increase the number of trades, the balance being determined by the hit probability model.
\newline
\newline
{\bf Multi-RfQ optimization:} In practice, as discussed in section \ref{GraphicalModel}, a dealer that wants to remain profitable needs to consider revenues of the full round-trip of buying and selling the bond. This translates into adding to the spread a compensation for the risk of holding the inventory until a matching opposite trade is closed. Such inventory risk complicates the problem in many ways: 1) there is uncertainty on when such RfQ will arrive, 2) other RfQs might arrive for this instrument in the same side or for different volumes, and the dealer might be interested in trading them in between. Therefore, the optimal pricing strategy for the current RfQ should incorporate these potential future scenarios in order to correctly price the risk. Such optimal pricing strategies require the use of stochastic optimal control theory; see \cite{avellaneda2008high, Gu_ant_2012} for a more detailed analysis. 

At the heart of these models is the hit probability model. The general formulation of the problem, though, does not have a closed-form solution unless specific limits are taken. For instance, let us discuss the first-order approximation in order arrival\footnote{Or equivalently, as shown in \cite{Gu_ant_2012}, the limit $t \rightarrow T$} derived by Avellaneda - Stoikov in their seminal paper \cite{avellaneda2008high}. Interestingly, this limit can be reproduced within our graphical model by considering a dealer that chooses the spread that maximizes revenues per trade using end of day market mid-prices to value the inventory. In terms of the exponential utility model, the optimization reads:
\begin{eqnarray}
\delta_{\text{opt}} = \text{argmax}_{\delta} \mathbb{E} \left[ -\exp(-\gamma R_T) |\text{do}(\delta), \text{RfQ}, \mathcal{Z}_t \right]
\end{eqnarray}
where we have assumed that the dealer had a position $q$ before receiving the RfQ, and no other RfQs are received in the interim. In this case, the revenues at the end of the day can be parametrized as a function of the spread charged and the mid-price movement until the market closes, namely:  $R_T = 1_\text{hit}(v\delta +(q + s v)(P_{m,T}-P_{m,t})) +(1-1_\text{hit}) q (P_{m,T} - P_{m,t})$. To compute this expected value is convenient to condition on the RfQ status ($RS$):
\begin{eqnarray}
\mathbb{E} \left[ -\exp(-\gamma R_T) |\text{do}(\delta), \text{RfQ}, \mathcal{Z}_t \right] = \nonumber \\  
\mathbb{E} \left[ -\exp(-\gamma\left(v\delta +(q + s v)(P_{m,T}-P_{m,t}\right)))|\text{do}(\delta), \text{hit}, \text{RfQ}, \mathcal{Z}_t \right] \times \nonumber \\ P(\text{hit} |\text{do}(\delta), \text{RfQ}, \mathcal{Z}_t)  +  \nonumber \\ 
\mathbb{E} \left[ -\exp(-\gamma q (P_{m,T} - P_{m,t}))|\text{do}(\delta), \text{no hit},  \text{RfQ}, \mathcal{Z}_t \right] \times \nonumber \\ \left(1-P(\text{hit} |\text{do}(\delta), \text{RfQ}, \mathcal{Z}_t)\right)
\label{eq10}
\end{eqnarray}
where the hit probability model naturally shows up again. Their estimation, as before, requires conditioning on at least $\sigma$, $BF$, $CF$, and $RF$. In this computation, though, the utility of revenues on hit is still a random variable which depends, according to our model, on the evolution of the mid-price. We model such dependence using a Brownian Motion model whose drift is contingent on the information asymmetry of the client requesting the RfQ. Such conditioning is also necessary in order to evaluate what we will refer to as {\it revenues on hit model} with historical data: as it will be detailed in the next section, to satisfy the back-door criterion and remove the do-operator from the expected value, the conditioning set must include at least the volatility $\sigma$, the RfQ features $RF$, and either the latent variable $IA$ or the drift $\mu$. Being latent variables, this necessarily means integrating over their distribution.

For simplicity, we choose ${\mathcal Z_t}$ to include $BF$ and $CF$ as well, since the extra variables don’t open new back-door spurious
paths for the revenues on hit, allowing us to use a single information set for both the hit probability model and the revenues on hit model. We can now proceed to evaluate the model. Let us first assume that the dealer does not expect informational advantage in this RfQ, and there is no other expectation of short-term directionality of the market. We can then use $P_{m,T} - P_{m,t} \sim N(0, \sigma^{2} (T-t))$ and compute the optimal spread:
\begin{eqnarray}
\delta_{\text{opt}}  = \gamma \sigma^2 (T-t) \left(s q + \frac{v}{2}\right) + \frac{1}{\gamma v}\log\left(1-\gamma v \frac{f(\delta_{\text{opt}})}{f'(\delta_{\text{opt}})}\right)
\end{eqnarray}
In this limit, the result reproduces the Avellaneda - Stoikov approximation \cite{avellaneda2008high}.

We can extend this model to the case in which the dealer also has a view on the informational advantage of the client. A simple model that captures the influence of the information asymmetry into the market dynamics is: $\hat{\mu} 1_{\text{IA} = 1}$. This means that if $\text{IA} = 1$, then $P_{m,T} - P_{m,t} \sim N(\hat{\mu} (T-t), \sigma^{2} (T-t))$. The conditioning set to compute revenues in this case does not change, as discussed in the next section. Working out again the utility maximization problem (Eq. \ref{eq10}),
the optimal spread has an extra additive correction that compensates for the information asymmetry risk:
\begin{eqnarray}
\delta_{\text{opt}}^\text{IA} =  \frac{1}{\gamma v} \log \left[\frac{p_\text{IA} e^{-\gamma (q+sv)\hat{\mu}(T-t)} + H (1 - p_\text{IA})}{p_\text{IA} e^{-\gamma q\hat{\mu}(T-t)} + H (1 - p_\text{IA})}\right]   
\end{eqnarray}
where $p_\text{IA} \equiv P(\text{IA}=1| CF)$. To arrive at this expression, we have additionally considered the following simple model, which captures the influence of the information asymmetry in the hit probability: $P(\text{hit}|\delta, \text{RfQ}, \text{IA} = 1, \mathcal{Z}_t ) = H P(\text{hit}|\delta, \text{RfQ}, \mathcal{Z}_t) = H f(\delta)$, with $H$ a constant in the range $0 \leq H \leq 1/ f(\delta)$.

The resulting spread correction is akin to the one analyzed by Glosten and Milgrom \cite{GLOSTEN198571} in a discrete time setup. As $p_\text{IA}$ increases, the dealer compensates the potential information asymmetry risk with a larger spread. In the limit $p_\text{IA} \rightarrow 1$, the client has to pay the full expected drift $\hat{\mu} (T-t)$. Interestingly, the required compensation decreases as $H$ increases, indicating that the hit probability model has correctly identified that the informed client is more likely to trade at the same spread than the uninformed one. This, in turn, reduces the need for an explicit spread protection term, $\delta_{\text{opt}}^\text{IA}$. 

In the limiting case of the multi-RfQ problem---equivalent to analyzing a single RfQ with a future liquidation time---inventory risk introduces an additive correction to the transactional risk term that is independent of the hit probability model. Intuitively, however, once we account for potential future RfQs, inventory risk becomes sensitive to the pricing strategy employed, which itself depends on the hit probability model to capture the client's responsiveness to quoted spreads. While a full causal analysis of the multi-RfQ setting lies beyond the scope of this work, we expect that the discussion surrounding the necessary conditioning set for estimating the hit probability and revenues-on-hit models remains applicable. In this multi-RfQ context, revenues are naturally linked across quotes, rather than isolated to single events.

This is particularly relevant because many derivations in the optimal market-making literature---see, for example, \cite{Gu_ant_2012, gueant2017optimal, closed_form_market_making}---assume simplified forms of the hit probability model, often modeling it as an exponential function of the spread:
\[
f(\delta) = p_0 \exp(-\alpha \delta),
\]
where $\alpha$ characterizes the client's price sensitivity. However, as we have shown, when these models are applied to optimal pricing---which constitutes a causal intervention---they can introduce bias if the estimation ignores spurious correlations present in historical data, which arise from the pricing strategies dealers have already employed. Thus, extending the causal inference framework to rigorously address the multi-RfQ problem remains an open and important challenge, warranting dedicated analysis.

\subsection*{RfQ revenue potential}

We define the RfQ revenue potential as:
\begin{equation}
P(R > 0|\text{do}(\delta), \text{RfQ}, \mathcal{Z}_t)
\end{equation}
The revenue of an RfQ can adopt any of the definitions discussed in previous sections. While it is a quantity closely related to price optimization, our focus here is on characterizing RfQs by their probability of being profitable—framing the problem as a classification task in Machine Learning terms. Revenue potential serves as a valuable indicator for dealers, enabling quick alerts in situations where the current pricing strategy is associated with a low likelihood of profitability, thereby suggesting a potential need for repricing.

To compute this expression, we use the sum and product probability rules to introduce the RfQ status variable ($RS$), and use the fact that $R > 0$ only when the dealer trades the RfQ, i.e. $\text{RS} = \text{hit}$:
\begin{equation}
P(R > 0|\text{do}(\delta), \text{RfQ}, \mathcal{Z}_t) = P(\text{hit}|\text{do}(\delta), \text{RfQ}, \mathcal{Z}_t) P(R > 0| \text{do}(\delta), \text{hit}, \text{RfQ}, \mathcal{Z}_t) 
\end{equation}
The first term is again the hit probability model, discussed in the previous section. The second term is closely related to the \textit{revenues-on-hit} model also introduced in the previous section, though it now appears framed as a classification problem: $P(R > 0| \text{do}(\delta), \text{hit}, \text{RfQ}, \mathcal{Z}_t)$. The discussion on how to estimate it from historical data applies equally in this context. As discussed in Section \ref{causal_analysis}, we need to include, at a minimum, the volatility $\sigma$, the RfQ features $RF$, and the information asymmetry state $IA$ in the conditioning set (or, alternatively to $IA$, the drift $\mu$). Since $IA$ is a latent variable, its inclusion requires integration over its distribution. With this conditioning set, the back-door criterion is satisfied, ensuring that the estimation specifically captures the direct effect of the spread on the potential revenues.

This conditioning set is smaller than the one required for the hit probability model. Therefore, we can either estimate the full model using the most restrictive (i.e., largest) conditioning set, or alternatively, integrate over the distribution of the variables that are not included but are still necessary for valid estimation. Let us focus here on the former case, which means adding $CF$ and $BF$ to the conditioning set:
\begin{eqnarray}
P(R > 0|\text{do}(\delta), \text{hit}, \text{RfQ}, \sigma, BF, CF, RF)) = \nonumber \\ \sum_{\text{IA}=0,1} P(R > 0| \delta, \text{hit}, \text{IA}, \sigma, RF) P(\text{IA} | \text{hit}, \text{RfQ},CF) 
\end{eqnarray}
Here, we have removed the do-operator on the right-hand side, as all back-door paths have been blocked, and we have applied the conditional independence structure of the graphical model to simplify the two probabilities involved~\cite{koller2009probabilistic, denev2015probabilistic}. Let us now use a metric that evaluates revenues at a later time $T>t$, i.e. $R_T =  v \delta_{t} + s v (P_{m,T} - P_{m,t})$. We also use the simple model introduced before for the influence of IA into the mid-price, i.e. $\mu(IA) = \hat{\mu} 1_{IA=1}$, and the Brownian motion for the price dynamics, $P_{m,T} = P_{m,t} + \hat{\mu} 1_{IA=1} (T-t) + \sigma \sqrt{T-t} Z$, with $Z\sim N(0,1)$. We can compute $P(R_T > 0| \delta_t, \text{hit}, \text{IA}, \sigma, RF)$ individually for each information asymmetry state, namely:
\begin{equation}
P(R_T > 0| \delta_t, \text{hit}, \text{IA},  \sigma, RF) = 
\begin{cases}
1-\Phi(\frac{-\delta_t - \hat{\mu} 1_{\text{IA} = 1} (T-t)}{\sigma \sqrt{T-t}})& \text{if } s = 1 \\
\Phi(\frac{\delta_t + \hat{\mu} 1_{\text{IA}=1} (T-t)}{\sigma \sqrt{T-t}}) & \text{if } s = -1
\end{cases}
\end{equation}
where $s$ is the side, and $\Phi(z) = P(Z \leq z)$ is the cumulative distribution of a standard normal random variable. 

\subsection*{Axe matcher}

When dealers hold excess exposure in certain bonds, they may be willing to offer more aggressive prices in order to liquidate those positions. In market jargon, such positions are referred to as \textit{axes}. A common question in this context is how to identify clients who are likely to be interested in trading on those axes. This is precisely the objective of the \textit{axe matcher} model, which, within the framework of our graphical model, can be formulated as follows:
\begin{equation}
    P(\text{hit}| \text{do}(\text{call} = 1), \text{do}(\delta), \text{axe} = 1, CF, BF, \mathcal{Z}_t)
    \label{eq_axe_matcher}
\end{equation} 
That is, we are assuming in this case that the dealer is performing interventions on two variables: 1) \textit{call}, indicating that a client is contacted by the dealer to raise interest in trading an axe; and 2) the spread $\delta$ quoted to maximize profits, conditional on the bond being axed. The model is, by definition, conditioned to bond and client features, since the axe refers to a specific bond and a specific client is called. Using common features to characterize both variables can improve overall estimation performance and help address gaps in historical data. Since the actual RfQ is not submitted prior to calling the client, the RfQ features $RF$ are not known in advance—except for the side, as axes are one-sided by construction. Therefore, the conditional distribution of $RF$ must be used, and integration must be performed over this distribution. It is important to note that, because the dealer is primarily interested in identifying clients for whom the commercial action (\textit{call}) has the greatest impact, the axe matcher model must, in practice, evaluate the \textit{average causal effect} (ACE)~\cite{pearl2016causal}:
\begin{eqnarray}
    \text{ACE} = P(\text{hit}| \text{do}(\text{call} = 1), \text{do}(\delta), \text{axe} = 1, CF, BF, \mathcal{Z}_t) - \nonumber \\ P(\text{hit}| \text{do}(\text{call} = 0), \text{do}(\delta), \text{axe} = 1, CF, BF, \mathcal{Z}_t)
\label{eq18}
\end{eqnarray}
This allows the dealer to avoid spending valuable commercial resources on clients who would likely trade the axe even without any intervention. In practice, both objectives must be balanced: selecting clients who exhibit both a sufficiently high {\it on-call hit probability} and a high average causal effect (ACE).

Let us focus on Eq. \ref{eq_axe_matcher}, since the discussion can be directly extended to the ACE. First, notice that according to our graphical model, we can decompose this probability in two parts: 1) the decision of the client to request a quote given the commercial action, and 2) the decision to trade when a price is quoted. Here, the channel of the RfQ could be a MD2C platform as before, or directly requested on a Single Dealer to Client (SD2C) platform, which in our model is equivalent to having a number of dealers $n = 0$. This allows us to write:
\begin{eqnarray}
    P(\text{hit}| \text{do}(\text{call} = 1), \text{do}(\delta), \text{axe} = 1, CF, BF,  \mathcal{Z}_t) = \nonumber \\  P(\text{hit}|  \text{do}(\text{call} = 1), \text{do}(\delta), \text{RfQ}, \text{axe} = 1, CF, BF, \mathcal{Z}_t)   \times \nonumber \\ P (\text{RfQ}| \text{do}(\text{call} = 1), \text{do}(\delta), \text{axe} = 1, CF, BF, \mathcal{Z}_t)
\end{eqnarray}
Factoring the axe matcher into these two probabilities provides a more tractable formulation. Since, in our model, there is no direct causal effect between the client's demand (captured by the $\text{RfQ}$ variable) and the spread $\delta$, assuming the existence of a conditioning set $\mathcal{Z}_t$ that blocks all the spurious correlations between these variables —without inducing new ones— allows us to simplify the second probability. A similar argument holds for the RfQ status ($RS$) with respect to the commercial action variable $call$ in the first probability. Under these assumptions, the overall axe matcher expression can be rewritten as:
\begin{eqnarray}
    P(\text{hit}| \text{do}(\text{call} = 1), \text{do}(\delta), \text{axe} = 1, CF, BF, \mathcal{Z}_t) = \nonumber \\P(\text{hit}| \text{do}(\delta), \text{RfQ}, \text{axe} = 1, CF, BF, \mathcal{Z}_t)  \times \nonumber \\ P (\text{RfQ}| \text{do}(\text{call} = 1), \text{axe} = 1, CF, BF, \mathcal{Z}_t)
\end{eqnarray}
Rewriting the second factor of Eq. \ref{eq18} in a similar manner, the ACE reads:
\begin{eqnarray}
    ACE = P(\text{hit}| \text{do}(\delta), \text{RfQ}, \text{axe} = 1, CF, BF, \mathcal{Z}_t) \times \nonumber \\ \left(P (\text{RfQ}| \text{do}(\text{call} = 1), \text{axe} = 1, CF, BF,  \mathcal{Z}_t) - \right. \nonumber \\ \left. P (\text{RfQ}| \text{do}(\text{call} = 0), \text{axe} = 1, CF, BF, \mathcal{Z}_t) \right)
\end{eqnarray}
The first term can be directly linked to the hit probability model, since conditioning on $\text{RfQ}$ allows us to remove any influence from the variable $axe$: at this point, the client's decision to trade is assumed to depend solely on the pricing, regardless of whether the dealer is axed or not. To estimate this model using historical data, as previously discussed, we must include the volatility $\sigma$ and the RfQ features $RF$—such as the requested volume and the number of competing dealers $n$—in the conditioning set. However, since these variables are not known in advance in the context of the axe matcher problem, we must integrate over their conditional probability distribution to properly account for their effect.

The second term is referred to in marketing literature as the \textit{uplift model}~\cite{katsov_algo_marketing}, and it enables us to isolate the causal effect of the intervention (i.e., the commercial action of calling the client) on the client's demand. From the perspective of causal inference, this quantity can be directly estimated from historical data: conditioning on $\text{axe}$ removes the spurious correlations between $\text{RfQ}$ and $\text{call}$. Therefore, in this case, the minimal conditioning set is empty, and we can write:
\begin{eqnarray}
    P (\text{RfQ}| \text{do}(\text{call}), \text{axe}, CF, BF) =  P (\text{RfQ}| \text{call}, \text{axe}, CF, BF)
\end{eqnarray}

These probabilities can be further simplified using the structure of the graphical model and Bayes' theorem:
\begin{eqnarray}
P (\text{RfQ}| \text{call}, \text{axe}, CF, BF) = \nonumber \\ = \frac{1}{Z} P(\text{call}|\text{axe}, \text{RfQ}) P(\text{axe}|\text{RfQ}) P(\text{RfQ}| CF, BF) 
\end{eqnarray}
where $Z$ is a normalization constant, and we have used conditional independence properties of the graph to write $P(\text{call}| \text{axe}, \text{RfQ}, CF, BF) = P(\text{call}|\text{axe}, \text{RfQ})$ and $ P(\text{axe}|\text{RfQ}, CF, BF) = P(\text{axe}|\text{RfQ})$. There are three relevant inferences in this model:
\begin{itemize}
\item The term $P(\text{call}|\text{axe}, \text{RfQ})$ which analyzes the proportion of RfQs on axes that happened after a commercial action (e.g. a call to the client). Therefore, this term captures the effectiveness of commercial campaigns in terms of conversions to RfQs. 
\item The term $P(\text{axe}|\text{RfQ})$ which estimates the distribution of axed RfQs.
\item The term $P(\text{RfQ}|CF, BF)$ estimates the distribution of RfQs with respect to client and bond features. This term captures the preferences of certain client segments for certain bonds. By projecting clients and bonds into common features can be estimated using techniques based on product-client matrix, like {\it collaborative filtering}, as shown by \cite{wrighta2023recommender} in the context of bond trading, or Latent Dirichlet Allocation, see \cite{hendricks_recommendation}. Since for a large number of bonds and clients, we expect to have sparse product-client matrix, these techniques are helpful to identify potential clients interested in trading bonds that did not necessarily trade them before in the dataset. 
\end{itemize}

\section{Causal analysis of the interventional probabilities}
\label{causal_analysis}

\subsection*{Hit probability model}

The hit probability model, as discussed in the previous section, is central to the business questions examined in this work. Our objective is to evaluate the causal quantity $P(RS = \text{hit} \mid \text{do}(\delta), \text{RfQ}, \mathcal{Z}_t)$. Specifically, we seek to determine the minimal set of variables in $\mathcal{Z}_t$ that enables estimation of this expression from historical (i.e., observational) data. According to Causal Inference theory~\cite{pearl2016causal}, a sufficient—though not necessary—condition is that the set $\mathcal{Z}_t$ satisfies the \textit{back-door criterion}. Given a causal graph and a pair of variables $X$ and $Y$, a set of variables $\mathcal{Z}$ satisfies this criterion if (i) no variable in $\mathcal{Z}$ is a descendant of $X$, and (ii) $\mathcal{Z}$ blocks all \textit{back-door paths} from $X$ to $Y$—that is, all paths that contain an arrow pointing into $X$. If such a set $\mathcal{Z}_t$ exists, we can validly equate the interventional and observational expressions:
\[
P(RS = \text{hit} \mid \text{do}(\delta), \text{RfQ}, \mathcal{Z}_t) = P(RS = \text{hit} \mid \delta, \text{RfQ}, \mathcal{Z}_t),
\]
allowing the right-hand side to be estimated using standard probabilistic methods. Analyzing the graphical model for the RfQ process, we see that there are two outgoing arrows from $\delta$, corresponding to two different paths in the graph: 
\begin{itemize}
\item The first path corresponds to the direct causal path to $RS$ under analysis.
\item The second path reaches $RS$ via the revenues $R$: $\delta \rightarrow R \leftarrow RS$. However, the arrow from $RS$ to $R$ reflects the reverse causal direction, meaning this path does not induce an additional causal effect of the spread $\delta$ on $RS$. In this structure, $R$ acts as a \textit{collider}\footnote{A variable is a {\it collider} on a path when both arrows along the path converge into it. Conditioning on a collider opens the path, potentially inducing spurious correlations between the variables that influence it.}, and unless $R$ is included in the conditioning set, the path remains blocked and does not introduce spurious associations between $\delta$ and $RS$. Therefore, we explicitly exclude $R$ from the conditioning set.
\end{itemize}
There are multiple back-door paths, though, that introduce spurious correlations via incoming arrows to the spread, unless properly blocked:
\begin{itemize}
\item The path via axe, which is a confounder\footnote{A variable functions as a {\it confounder} along a path when it emits arrows to both variables on that path. If such confounders are not included in the conditioning set, they can induce spurious correlations between the variables they influence.}: $\delta \leftarrow axe \rightarrow RfQ \rightarrow RS$. This path is blocked since we are conditioning on $\text{RfQ}$, i.e. on having received an RfQ for which the dealer will provide a quote.
\item The paths involving the client features $CF$ and bond features $BF$ act as sources of confounding between $\delta$ and $RS$ through different mechanisms. Some paths are mediated by the variable $\text{RfQ}$, but their spurious influence is already blocked by the conditioning on $\text{RfQ}$. Other paths are mediated by latent variables such as the client reservation spread $\delta_\text{res}$, the spreads offered by other dealers $\delta_{\text{dealer}}$, the price discovery variable $PD$, and the information asymmetry state $IA$. An example of these paths is: $\delta \leftarrow CF \rightarrow \delta_{\text{dealer}} \rightarrow RS$. We can block the influence of these paths by conditioning on the observable variables $BF$ and $CF$. While conditioning on the mediators themselves\footnote{A variable is a \textit{mediator} if it lies on a directed path from the treatment to the outcome, transmitting part of the causal effect.} would achieve a similar blocking effect, these variables are latent and thus cannot be directly observed. To include them in $\mathcal{Z}_t$, we would need to integrate over their distribution—at the cost of introducing additional model and numerical complexity.
\item The paths introduced by the volatility $\sigma$, for example the ones mediated by the dealer's prices, e.g. $\delta \leftarrow \sigma \rightarrow \delta_{\text{dealer}} \rightarrow RS$, or the client's reservation price, for example $\delta \leftarrow \sigma \rightarrow \delta_{\text{res}} \rightarrow RS$. Since $\sigma$ is a confounder in all these paths, we block them by including the volatility in the conditioning set. The paths that pass via the revenues $R$ are closed as far as this variable is not included in the conditioning set, since it acts again as a collider in these paths. 
\item The paths via the RfQ features $RF$. On the one hand, the path  $\delta \leftarrow RF \rightarrow R \leftarrow RS$ is blocked since $R$, the revenues, is a collider with respect to any direct or indirect path connecting it to $RS$. By not conditioning on revenues, these paths are blocked. Paths passing through $BF$, $CF$ and $\text{RfQ}$ are blocked due to the conditioning on those variables. Finally we have paths connecting RfQ features to $RS$ via latent variables like $\delta_\text{res}$ and $\delta_\text{dealer}$, which are open. Again, to block these spurious influences using observable variables we can include $RF$ in the conditioning set. 
\end{itemize}

In conclusion, the minimal set of variables required for valid estimation of the hit probability model, according to our graphical model, consists of the volatility $\sigma$, the RfQ features $RF$, the bond features $BF$, and the client features $CF$, in addition to $\text{RfQ}$, which is inherently part of the model’s definition. If dealers choose to implement pricing policies based on a reduced set of features, they must still compute conditional probabilities using this minimal set. This, in turn, requires integrating over the distribution of the excluded variables:
\begin{eqnarray}
P(\text{hit} |\text{do}(\delta), \text{RfQ}) = \sum_{RF} \sum_{BF} \sum_{CF} \int P(\text{hit} |\delta, \text{RfQ}, \sigma, RF, BF, CF)  \nonumber \\ \times P(RF, BF, CF | \text{RfQ}) dP(\sigma)
\end{eqnarray}
where we have used the conditional independence structure of the graphical model to simplify the probabilities \cite{koller2009probabilistic, denev2015probabilistic}. 

\subsection*{Revenues on hit model}

The {\it revenues on hit} model makes inferences on revenues conditional to wining the RfQ. We have seen two related inferences in the previous section, one estimating expected revenues and the other computing the revenue potential, expressed as a binary condition $R > 0$. In either case, our goal is again to find the minimum set of variables in $\mathcal{Z}_t$ that satisfies the back-door criterion, allowing us to estimate this probability using historical data. By inspecting the graphical model, we identify the following causal paths linking revenues $R$ with spread $\delta$:
\begin{itemize}
\item The direct causal path $\delta \rightarrow R$, as revenues are directly linked to the spread charged with respect to the mid-price.
\item The indirect causal path mediated by RfQ status ($RS$): $\delta \rightarrow RS \rightarrow R$, which is blocked since the revenue potential model is conditioned to $RS$ by definition. 
\end{itemize}

Let us now analyze the back-door paths that introduce spurious correlations, i.e. they have arrows pointing to the intervened variable, the spread $\delta$:
\begin{itemize}
\item  Paths that pass via $RS$ and $\text{RfQ}$, or latent features like $\delta_\text{dealer}$, $\delta_\text{res}$, $PD$ and $IA$. They are blocked by the conditioning on $RS$.
\item The path via the RfQ features $RF$: $\delta \leftarrow RF \rightarrow R$, where $RF$ acts as a confounder. $RF$ needs to be included in the conditioning set $\mathcal{Z}_t$ in order to block this back-door path. Conditioning on $RF$ also blocks other back-door paths that connect $\delta$ to $RF$ via latent features like $\delta_\text{dealer}$ and $\delta_\text{res}$.
\item Similarly, the back-door path created by the volatility $\sigma$, a confounder, can be blocked by including $\sigma$ in the conditioning set. 
\item The back-door paths mediated by the drift $\mu$ through the information asymmetry variable $IA$ are particularly relevant, as they represent mechanisms by which revenues may be influenced by clients' informational advantage over dealers. These paths suggest that dealers might be adjusting their spreads to account for clients who historically exhibit signs of informational superiority—that is, clients for whom the market tends to move favorably after execution in ways that cannot be explained purely by chance. Some of these paths, such as $\delta \leftarrow CF \rightarrow IA \rightarrow \mu \rightarrow R$, can be blocked by conditioning on observable client or bond features. However, other paths—such as $\delta \rightarrow RS \leftarrow \delta_\text{res} \leftarrow IA \rightarrow \mu \rightarrow R$—cannot be blocked in this way. In such cases, conditioning on $RS$, which acts as a collider, introduces spurious correlations between $\delta$ and $R$. Since $RS$ is part of the model’s definition, this collider path cannot be avoided unless we condition on either $IA$ or $\mu$. Both $IA$ and $\mu$ are latent variables, so conditioning on them requires integrating over their distributions in order to apply the back-door criterion. Importantly, such conditioning also blocks the other paths, eliminating the need to include client or bond features in the minimal conditioning set. In practice, estimating $\mu$ in the presence of information asymmetry typically requires separate models for each state of $IA$, making $IA$ the more natural choice for conditioning in the estimation of the model.
\end{itemize}
The conclusion, in this case, is that we cannot estimate the revenues-on-hit model using only observational variables such as $\sigma$ and $RF$. However, if the model explicitly includes the information asymmetry state $IA$, we are permitted to remove the $\text{do}$-operator and compute the relevant probabilities using standard probabilistic methods, namely:
\begin{align}
 P(R >0 | \text{do}(\delta), \text{hit}, \text{RfQ}, \sigma, RF) = \sum_{\text{IA}= 0,1} &\; P(R >0 | \delta, \text{hit}, IA, \sigma, RF) \nonumber \\
&\times P(IA|\delta, \text{hit}, \text{RfQ}, \sigma, RF)
\end{align}
where again, we have used the conditional independence properties of the graph to slightly simplify the probabilities in the right hand side. 

\subsection*{Uplift model}

The uplift model in the context of the axe matcher quantifies the effect on client demand of a commercial action that recommends trading on axes, compared to clients who trade exogenously. To estimate this model, we must compute probabilities of the form
\[
P(\text{RfQ} \mid \text{do}(\text{call} = c), \text{axe} = 1, CF, BF, \mathcal{Z}_t),
\]
where $c \in \{0,1\}$. We rely on the causal graphical model to determine under what conditions this quantity can be estimated from historical data. In this case, the analysis is straightforward: all back-door paths between $\text{call}$ and $\text{RfQ}$ are mediated by the variable $\text{axe}$, which acts as a confounder. Conditioning on $\text{axe}$ is sufficient to block these paths. Therefore, we can safely choose the empty set for $\mathcal{Z}_t$, and write:
\begin{equation}
P (\text{RfQ}| \text{do}(\text{call} = c), \text{axe} = 1, CF, BF) =  P (\text{RfQ}| \text{call} = c, \text{axe} = 1, CF, BF)
\end{equation}

\section{Model specification}

So far we have used Causal Inference theory to express business causal interventions in terms of probabilities that can be estimated using historical data. When it comes to actually computing those probabilities, we need to provide a model specification. Here, two main routes can be taken: 
\begin{itemize}
\item Propose a generative model \cite{murphy2013machine}, i.e. specify the joint distribution of the random variables in the graph, which can be factored into conditional distributions of each node with respect to their parents nodes \cite{koller2009probabilistic, denev2015probabilistic}. The model is then estimated using standard techniques like Maximum Likelihood Estimation (MLE), Maximum a Posteriori (MAP) or a full Bayesian approach where we compute the posterior distribution of the parameters conditional to the historical data. 
\item Propose specific discriminative models for each of the conditional distributions that need to be computed, and fit them independently to the data.
\end{itemize}

Generative models offer the advantage of parsimony, as they aim to model the joint distribution of all relevant variables. This allows them to capture specific mechanisms underlying the data generation process, and, once estimated, they can be used to compute any conditional probability through standard probabilistic rules. In contrast, discriminative models estimate conditional distributions directly, often without assuming a common underlying generative structure. Sophisticated Machine Learning techniques~\cite{bishop2007, murphy2013machine} can be applied to enhance their predictive power. However, these models are typically more opaque in their internal workings and may struggle to represent the structural mechanisms behind data generation. Nevertheless, because they are trained specifically to approximate conditional distributions, discriminative models often outperform generative models in predictive tasks. Moreover, they benefit from advances in modern Machine Learning, including regularization techniques and ensemble methods, which help mitigate overfitting and improve out-of-sample generalization.

\subsection*{A generative model of the RfQ process}

We propose a generative model for the RfQ process based on the graphical model discussed in previous sections. The model builds upon the assumptions from \cite{fermanian2016behavior}, extending it to add extra mechanisms not included in this seminal work. We will focus on modeling the relevant conditional distributions required to compute the causal interventions discussed in this article:
\begin{itemize}
\item As it is usual in the optimal market-making literature \cite{gueant2016financial, cartea2015algorithmic}, we model the arrival of RfQs as a Poisson process $N_t$ with intensity $\lambda$. In particular, of interest for our model is the conditional distribution of arrival of an RfQ in a differential of time $[t, t + dt]$, which is the model we used for the binary variable $\text{RfQ}$ in the graphical model, conditional to the parent nodes, in particular the RfQ features which include the time $t$ of the RfQ:
\begin{eqnarray}
P(\text{RfQ} = 1| BF, CF, RF, PD, IA, \text{call}, \text{axe}) = \nonumber \\ P(N_{t+dt} - N_t = 1| BF, CF, RF, PD, IA, \text{call}, \text{axe}) = \nonumber \\
\mathbb{E} [\lambda| BF, CF, RF, PD, IA, \text{call}, \text{axe}] dt
\end{eqnarray}
We now use a linear model for the conditional intensity:
\begin{eqnarray}
\mathbb{E} [\lambda | BF, CF, RF, \sigma, IA, \text{call}, \text{axe}) ] = \nonumber \\ \lambda_0 + \sum_i \lambda_{c, i} CF_i + \sum_j \lambda_{b,j} BF_j + \sum_k \lambda_{r,k} RF_k + \lambda_{PD} \mathbbm{1}_{PD = 1} + \lambda_{IA} \mathbbm{1}_{IA = 1}  \nonumber \\ + \lambda_{\text{call}} \mathbbm{1}_{\text{call} = 1} + \lambda_{\text{axe}} \mathbbm{1}_{\text{axe} = 1} 
\end{eqnarray}
\item The conditional distribution of $\text{call}$ with respect to its parent node, $\text{axe}$, is modelled using a Bernoulli distribution: 
\begin{equation}
P(\text{call} = 1 | \text{axe} = a) = p_a
\end{equation}
where $a = \{0,1\}$.
\item As in \cite{fermanian2016behavior}, we model the reservation spread of the client using a Gaussian distribution on the normalized half-spread: $\delta_\text{res} / \Delta$, where $\Delta$ reflects market liquidity conditions, using for instance half the Composite Bloomberg Bond Trader (CBBT) bid-ask spread or any other relevant benchmark spread. Since bond, client and RfQ features, as well as volatility $\sigma$ and information asymmetry $IA$, are parent nodes of the reservation price, we need to model the conditional distribution, for which a linear regression model is used, namely: 
\begin{eqnarray}
P(\delta_\text{res}|\sigma, CF, BF, RF, IA) = \nonumber \\ \mathcal{N}(\frac{\delta_\text{res}}{\Delta}| a_{res} + b_{res} \sigma + \sum_i c_{\text{res}, i} CF_i + \sum_j d_{\text{res}, j} BF_j + \sum_k e_{\text{res},k} RF_k + \nonumber \\ f_\text{res} \mathbbm{1}_{IA = 1}, \sigma_\text{res}^2)
\end{eqnarray}
where $\sigma_\text{res}$ is the standard deviation of the distribution of client reservation prices. 
\item For the conditional distribution of the spreads quoted by other dealers, we adopt the Skew Exponential Power (SEP) distribution, as proposed in~\cite{fermanian2016behavior}, applied to the normalized dealer half-spread ($\delta_{\text{dealer}} / \Delta$). To visually assess the appropriateness of this choice, we leverage a structural property of our model: the conditional independence of dealer quotes given their parent features. Under this assumption, the cover price—defined as $\delta_\text{cover} = \max\{\delta_{\text{dealer}, 1}, \dots, \delta_{\text{dealer}, n}\}$ for buy-side RfQs (and the minimum for sell-side RfQs)—follows the distribution of the maximum of $n$ independent random variables, each drawn from the same SEP distribution conditional on parent features. Fig~\ref{fig2} displays this distributional fit using the SEP assumption. Notably, despite our dataset differing from the one used in~\cite{fermanian2016behavior}, the results offer further evidence of the robustness of the SEP specification. To model the conditional distribution of dealer spreads with respect to their parent nodes—which include bond, client, and RfQ features, as well as volatility—we use a linear model:
\begin{eqnarray}
P(\delta_\text{dealer}|\sigma, CF, BF, RF) =  \nonumber \\ \text{F}_{\theta_1, \theta_2, \theta_3, \theta_4}(\frac{\delta_\text{dealer}}{\Delta}| a_{\text{d}} + b_{\text{d}} \sigma + \sum_i c_{\text{d}, i} CF_i + \sum_j d_{\text{d}, j} BF_j + \sum_k e_{\text{d},k} RF_k )
\end{eqnarray}
where $\theta_1$ determines the location, $\theta_2$ the scale, $\theta_3$ the shape and $\theta_4$ the asymmetry.

\begin{figure}[H]
\centering
\includegraphics[scale=0.5]{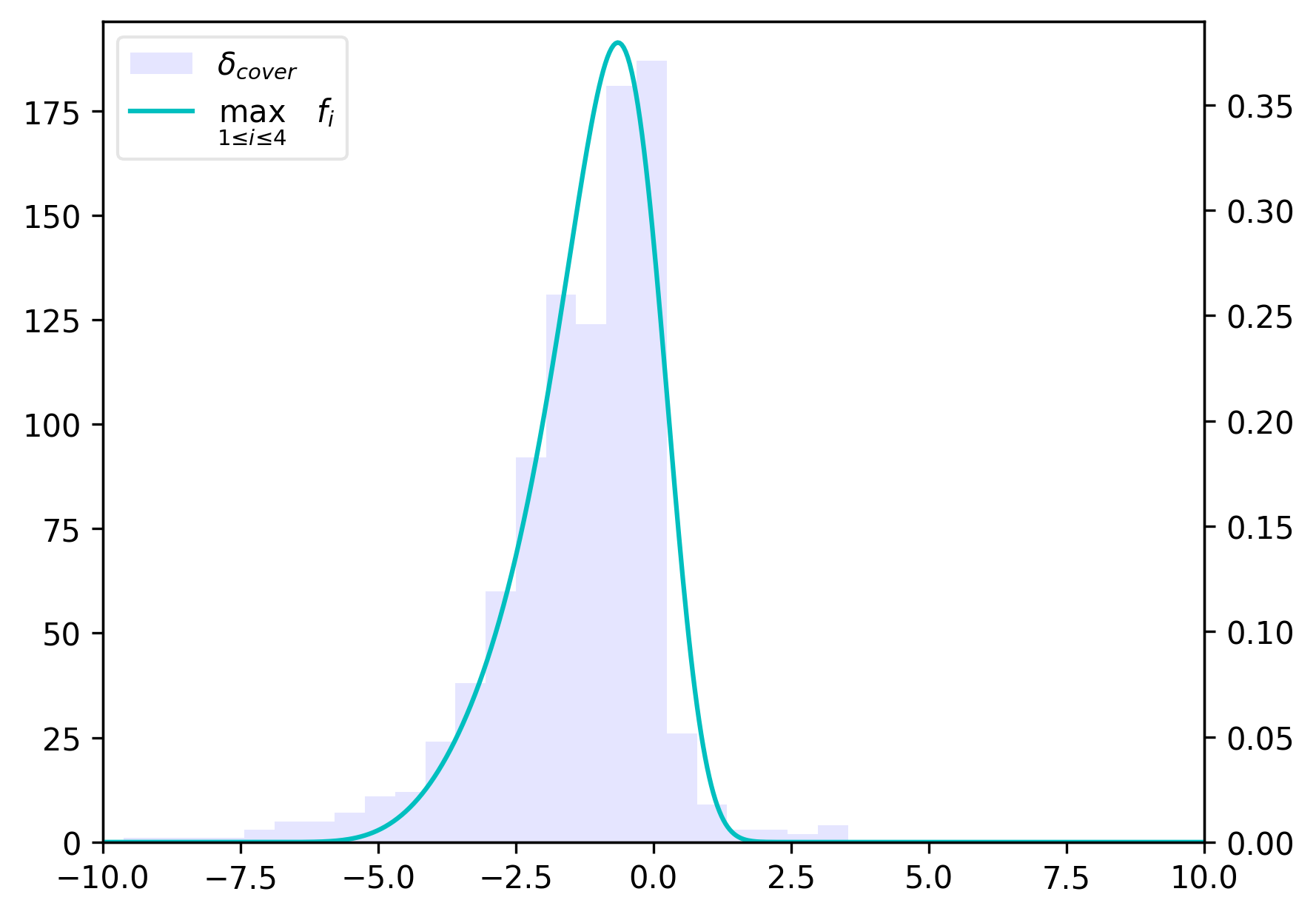}
\caption{Probability density function of the best quotes from other four dealers, as implied by the model, plotted against the empirical histogram of cover prices in buy RfQs.
 \label{fig2}}
\end{figure}

\item For the distribution of the price discovery variable $PD$, conditional on its parent variables $BF$ and $CF$, one could, in principle, use Bernoulli distributions. However, this approach risks an exponential increase in the number of parameters if a large number of features is included. To address this issue while preserving interpretability and tractability, we model $PD$—a binary random variable—using a logistic regression model, namely:
\begin{equation}
P(PD = 1|BF, CF) = \frac{1}{1 + \exp\left({a_{\text{p}} + \sum_i b_{\text{p}, i} CF_i + \sum_j c_{\text{p}, j} BF_j  }\right)}
\end{equation}
\item A similar approach can be followed to model the information asymmetry of the client $IA$, whose distribution conditional to client features $CF$ can also be described using a Logistic regression.
\item Given the previous assumptions, the distribution of the RfQ status variable ($RS$) is deterministic conditional on its parent nodes. These include: $\text{RfQ}$, the request-for-quote event; $\delta_\text{res}$, the client's reservation price; $\delta_\text{dealer}$, the spreads quoted by competing dealers; $\delta$, the spread quoted by the dealer; and $PD$, the price discovery indicator. The discussion below centers of the case of buy RfQs, but analogous expressions apply to the case of sell RfQs:
\begin{itemize}
\item $RS = \text{hit}$: A trade occurs when a client requests a quote ($\text{RfQ} = 1$), is interested in trading ($PD = 0$), the dealer's quoted spread is better than those of competitors (i.e., $\delta \geq \delta_\text{best} \equiv \max \{ \delta_{\text{dealer}, 1}, \dots, \delta_{\text{dealer}, n} \}$), and the quoted spread is no worse than the client's reservation price ($\delta \geq \delta_\text{res}$). Within the model, we assume that tied RfQs—cases where the dealer's quote matches the best competitor—are won by default. However, a random allocation mechanism could be introduced to account for tie-breaking behavior if desired.
\item $RS = \text{missed}$: A missed trade occurs when a client requests a quote ($\text{RfQ} = 1$), is interested in trading ($PD = 0$), the dealer's quoted spread is worse than the best competing quote (i.e., $\delta < \delta_\text{best}$), and the best competitor's quote is still acceptable to the client (i.e., $\delta_\text{best} \geq \delta_\text{res}$). Among missed RfQs, it is possible to further distinguish the status $\text{cover}$—if this information is provided by the platform—which occurs when the dealer's quote is the second best among all quotes submitted.
\item $RS = \text{passed}$: There are two main scenarios that lead to a missed trade. In the first, a client requests a quote ($\text{RfQ} = 1$) but is not actually interested in trading ($PD = 1$), using the RfQ process instead as a means of price discovery. In the second scenario, the client is interested in trading ($PD = 0$) and requests a quote ($\text{RfQ} = 1$), but the client's reservation spread is more conservative than any of the quotes received —that is, $\delta_\text{res} > \max \{ \delta, \delta_\text{best} \}$.
\end{itemize}
\item As we discussed in the context of optimal pricing with information asymmetry, a simple model for the conditional distribution of the drift $\mu$ of the market after closing the RfQ is $\mu = \hat{\mu} \mathbbm{1}_{IA = 1}$, where $\hat{\mu}$ is a parameter of the model.
\item Finally, the distribution of the revenue $R$ depends on the definition of revenues discussed in previous sections. If we use the short-term flow value definition of revenue we have
\begin{equation}
R_{t+h} = 1_\text{hit} \left(v \delta_{t} + s v (P_{m,t + h} - P_{m,t}) \right)= 1_\text{hit} \left(v \delta_{t} + s v \mu h + sv \sigma (W_{t+h} - W_t)\right)
\end{equation}
where $h > 0$ is the time window, and $\text{MXF}$, the market external factors, are modeled using a Wiener process $\text{MXF} = W_{t+h} - W_t \sim \mathcal{N}(0, h)$. 
\end{itemize}
An interesting effect discussed in~\cite{fermanian2016behavior} that we have incorporated in our framework is the \textit{partial participation effect}, which reflects the empirical observation that not all dealers invited to an RfQ actually respond with a quote. We model dealer participation by assuming that each dealer independently chooses to quote with a fixed probability \( p_{\text{quote}} \), which is identical across all dealers. Although this mechanism is not explicitly depicted in the graphical model, it can be readily incorporated by adding an additional parent node to \( \delta_{\text{dealer}} \), representing the dealer’s quoting decision and assumed to be independent of other factors. As a result, the number of dealers who effectively submit quotes for a given RfQ follows a Binomial distribution.

\subsection*{Discriminative models of the RfQ process}

When modeling the conditional distributions required to estimate the effects of causal interventions, we can leverage the extensive corpus of Machine Learning (ML) models available in the literature. These models offer flexible tools for specifying conditional distributions and fitting parameters from data. Within the standard model selection framework, this process typically involves setting aside a validation dataset for comparing alternative models and selecting the one that yields the best out-of-sample performance according to predefined quality metrics for the predictive task at hand. However, several important considerations arise when applying this framework to the RfQ process, which may influence both the choice of models and the interpretation of their outputs:
\begin{itemize}
\item Discriminative Machine Learning (ML) models must be trained on observable features. In the RfQ process, however, certain features such as the cover spread are only available when the RfQ is traded by the dealer and only after the negotiation has concluded. Even when post-trade reporting datasets are available, the relevant information (e.g., the final traded spread) becomes accessible only after the fact. As such, these features are not available at inference time. This creates a challenge, as some features are only partially observable. One option is to exclude these features entirely from both training and inference, accepting the loss of potentially valuable information. Alternatively, one may apply data imputation techniques~\cite{amen_2022}, or restrict the selection of models to those capable of handling partially observed inputs.
\item Discriminative Machine Learning (ML) models typically do not incorporate structural mechanisms of the RfQ process that constrain the set of feasible outcomes. For instance, we expect that rational investors will be less likely to trade when offered worse spreads, \textit{ceteris paribus}. However, such monotonic behavior is not inherently enforced by most ML models~\cite{dixon2020machine}. In contrast, the generative model described in the previous section naturally encodes this behavior through the introduction of the client's reservation spread, which acts as a structural threshold for trade execution.
\end{itemize}

\section{Generative versus discriminative models for causal interventions: empirical results}

To further analyze these points, we now turn our attention to the specification of the hit probability model, given by
\[
P(\text{hit} \mid \delta, \text{RfQ}, \sigma, RF, BF, CF),
\]
which, as shown in the previous section, lies at the core of all causal interventions considered in this work.

Using the generative model, we can compute this predictive probability distribution. For a bid, for example, this reads:

\begin{eqnarray}
P(\text{hit} \mid \delta, \text{RfQ}, \sigma, RF, BF, CF) = 
\left(1 - P(PD=1 \mid BF, CF)\right) \times \nonumber \\
\sum_{a = 0}^{1} P(IA = a \mid CF) \int d\delta_\text{res} \, f_\text{res}(\delta \mid \sigma, RF, BF, CF, a) \times \nonumber \\
\sum_{k=0}^{n} \binom{n}{k} p_{\text{quote}}^k (1 - p_{\text{quote}})^{n - k} \times \nonumber \\ \int \cdots \int 
\left( \prod_{i=1}^k d\delta_{\text{dealer}, i} \, f_\text{dealer}(\delta_{\text{dealer}, i} \mid \sigma, RF, BF, CF) \right) \times \nonumber \\
\mathbbm{1}_{\delta \geq \max \{ \delta_{\text{dealer}, 1}, \ldots, \delta_{\text{dealer}, k} \}}
\end{eqnarray}

As previously discussed, the parameters of the distributions in this model can be estimated using standard probabilistic methods. For instance, \cite{fermanian2016behavior} employs Maximum Likelihood Estimation (MLE) to fit a related version of the hit probability model over a dataset of RfQs\footnote{In \cite{fermanian2016behavior}, the authors focus specifically on the hit probability model, without incorporating mechanisms to account for information asymmetry or price discovery.}. One advantage of the generative approach is its ability to handle a more granular categorization of RfQ outcomes, enabling the calculation of the likelihood for each outcome using the structural specification described in the previous section. However, this estimation approach comes with the drawback of requiring a custom implementation of the integrals over latent variables—both in the likelihood function and the predictive distribution. To mitigate this complexity, one can resort to the Expectation-Maximization (EM) algorithm~\cite{murphy2013machine}, which reduces the computational burden of integrating over latent variables. A more detailed discussion of this approach in the context of the RfQ process will be included in \cite{paloma_thesis}.

We evaluate two discriminative models: a logistic regression and a tree-based model, the Light Gradient Boosting Machine (LightGBM). Logistic regression serves as a simple benchmark that inherently respects monotonicity constraints on the predicted probability with respect to input features—particularly the quoted spread. In contrast, LightGBM is a powerful, widely used model known for its strong out-of-the-box predictive performance. Although it does not enforce monotonicity, we employ it to explore the predictive frontier of the problem, assessing the potential gains in accuracy when the monotonicity constraint is relaxed. Both of them are efficient and interpretable, making them standard candidates to implement the hit probability model ~\cite{zhou2024explainable}. A promising alternative involves neural network architectures that explicitly incorporate monotonicity constraints; see~\cite{dixon2020machine} for a related application in the context of derivatives pricing. These models aim to combine the principled structure of monotonic estimators with the flexibility and representational power of neural networks. Results from this line of research will be presented in future work~\cite{paloma_thesis}.

To compare the generative model against its discriminative alternatives, we focus on a simplified version of the problem, following a setup similar to that of~\cite{fermanian2016behavior}:
\begin{itemize}
\item For simplicity, we assume that price discovery is not the motivation in the passed RfQs, but they correspond to situations where none of the quoted prices—either from the dealer under analysis or the competitors—fell within the client’s reservation price. This allows us to fix $PD = 0$ across the entire dataset, effectively treating all RfQs as trade-intended and simplifying the modeling assumptions. The task remains framed as a binary classification problem: predicting whether a given RfQ results in won (hit) or lost (missed or passed). This formulation aligns with practical dealer needs, as it enables estimation of trading probability without conditioning on observing actual transactions.
\item We assume that clients in the dataset do not possess information asymmetry, i.e., we set $IA = 0$. While this assumption may not always hold in practice and may degrade the model's predictive accuracy, it introduces the same structural bias across both the generative and discriminative models. As such, it remains valid for benchmarking purposes.
\end{itemize}

The choice of features is critical to ensure that the results are not driven by spurious correlations. We guide our feature selection based on the findings from previous sections regarding the minimal conditioning set that satisfies the back-door criterion. To identify a relevant subset of variables for model estimation, we combine domain-specific business knowledge with a standard feature selection technique: the relative feature importance derived from a Random Forest Classifier applied to the training set (see details below). The final set of selected features can be grouped into three categories. First, we consider the variables that are explicitly modeled as nodes in the graphical model and are present in the minimal conditioning set:
\begin{itemize}
\item The \textit{normalized half-spread} ($\delta / \Delta$), introduced previously, which adjusts the dealer's quoted spread $\delta$ by a market liquidity proxy—such as half the bid-ask spread of a composite benchmark like CBBT.
\item The \textit{yield volatility}, represented as $\sigma$ in the graphical model, which is calculated as the standard deviation of daily bond yield changes over the past 30 days.
\end{itemize}
Next, our algorithm has selected the following features that are related to the characteristics and market activity of the bond itself, corresponding to the set of variables labeled as $BF$ in our graphical model:
\begin{itemize}
\item The \textit{bond DV01}, representing the bond’s sensitivity to interest rate changes, measured as the price change corresponding to a 1 basis point shift in yield.
\item The \textit{average days between buy-side RfQs (per ISIN)}, calculated as the average time between consecutive buy-side RfQs for the same ISIN over the past 15 days.
\item The \textit{average days between sell-side RfQs (per ISIN)}, defined analogously to the previous feature, but for sell-side RfQs.
\item The \textit{average number of dealers per ISIN}, computed as the 15-day average number of dealers quoting RfQs for a given ISIN.
\item The number of \textit{days to maturity} of the bond.
\end{itemize}
Finally, we include features specific to the RfQ level, corresponding to the variable $RF$ in the graphical model:
\begin{itemize}
\item The \textit{number of dealers in competition} for the RfQ.
\item The \textit{DV01 exposure of the RfQ}, which scales the bond DV01 by the notional volume $v$ of the RfQ, providing a risk-adjusted measure of the quote size.
\end{itemize}

In our dataset, no client-related variable —corresponding to the node $CF$ in the graphical model—exhibited significant predictive power. We attribute this outcome to the fact that the analysis is confined to a specific segment of high-tier institutional clients, whose behavior may be relatively homogeneous. It is also worth noting that we experimented with incorporating past cover price information through engineered features, such as rolling averages of the distance between the dealer's quoted price and the observed cover price. However, these features did not lead to significant improvements in predictive performance. Developing more effective approaches to embed cover price information into discriminative models through feature engineering remains an open area of research. Additionally, we emphasize that the relevance and impact of specific features may vary across datasets. As such, feature selection should be adapted to the characteristics of the data at hand and guided by both domain knowledge and empirical validation.

We use a proprietary dataset of RfQs from BBVA, consisting of 102{,}437 RfQs, of which 5{,}738 are labeled as hits. More details about the data are available in \nameref{S1_Table}. The dataset is split into training, validation, and test sets, with the test set reserved for final model evaluation. The validation set is used for model selection and hyperparameter tuning in the discriminative ML models. Full details of the preprocessing steps are provided in \nameref{S1_Appendix}. For the logistic regression model, we tune the inverse regularization strength, setting it to 100 with an L2 penalty. For the LightGBM model, the best-performing hyperparameters identified are: column sample by tree = 1, learning rate = 0.01, minimum child samples = 50, number of estimators = 500, number of leaves = 15, and subsample = 0.6. The generative model does not require hyperparameter tuning and is trained using Maximum Likelihood Estimation (MLE) on the training set.  Given the class imbalance in the dataset, we apply class weighting during training for both the generative and discriminative models. In addition, standard outlier cleaning techniques are applied to preprocess the dataset.

To evaluate the performance of our models, we employ two complementary metrics: the Area Under the Receiver Operating Characteristic Curve (AUC-ROC) and the Balanced Brier Skill Score (BBSS), based on the Brier score~\cite{brie_1950_verification}. The AUC-ROC is a widely used metric for classification tasks, particularly valuable when dealing with imbalanced datasets. It measures the model’s ability to distinguish between classes across various threshold settings, providing a summary of the trade-off between true positive and false positive rates. Due to its robustness in skewed class distributions, we adopt the AUC as our primary metric for model selection.

We used the Balanced Brier Skill Score to assess the calibration and overall accuracy of probabilistic predictions. Unlike the AUC, which evaluates the model’s ranking capability, the Brier score measures the mean squared error between the predicted probabilities and the actual binary outcomes. The Balanced Brier Score (BBS) adjusts this metric by using the empirical frequency $w_m$ of the majority class (lost RfQs) in the training set, namely:
\[
\text{BSS} = \frac{\sum_{i=1}^{N} w_m^{1-o_i} (1-w_m)^{o_i}(f_i - o_i)^2}{\sum_{i=1}^N w_m^{1-o_i}(1-w_m)^{o_i}}
\]
where \( f_i \) is the predicted probability for instance \( i \), and \( o_i \in \{0, 1\} \) is the observed outcome. The Balanced Brier Skill Score is then defined as $\text{BBSS} = 1 - \frac{\text{BBS}}{\text{BBS}_{\text{bench}}}$, where $\text{BBS}_{\text{bench}}$ is the BBS of a simple majority class benchmark, where we predict a constant probability for all instances, equal to $w_m$. Notice that $\text{BBS}_\text{bench}$ can be computed analytically:

\[
\text{BBS}_\text{bench} = \frac{1}{2} - w_m (1-w_m)
\]

A worst-case Balanced Brier Score (BBS) is $0.5$, while a perfect model achieves a score of $0$. In terms of skill, the Balanced Brier Skill Score (BBSS) equals $0$ when the model's performance matches that of the benchmark, and reaches $1$ for a perfect model. This metric provides a succinct summary of the information conveyed by calibration plots, which group predicted probabilities into bins and compare the average predicted probability within each bin to the observed frequency of positive outcomes. Such plots—and, by extension, the Brier score—are particularly informative in settings where the predicted probabilities themselves, rather than just binary class labels, directly influence decision-making, as is the case in the business problems analyzed in this work.

The results for AUC-ROC and BBSS in the test set for the three models considered as well as the majority class benchmark are shown in Table \ref{tab1}.

\begin{table}[H]
\centering
 \begin{tabular}{| r || c | c | c |}
	\hline
    {\bf Model} & {\bf AUC-ROC} & {\bf BBSS} \\\hline
	Generative & 0.742 & 0.238 \\\hline
    Logistic regression & 0.684 & 0.446 \\\hline
    LightGBM & 0.743 & 0.406 \\ \hline
    Majority Class & 0.5 & 0.0 \\ \hline
  \end{tabular}
  \caption{Predictive scores: AUC-ROC and BBSS for the three models analyzed. We also show the metrics for the benchmark model, the majority class. Both the generative model and LightGBM achieve comparable predictive performance in terms of ROC-AUC, outperforming logistic regression. While the generative model exhibits lower calibration performance, as indicated by a reduced BBSS, it uniquely satisfies the key business constraint of spread monotonicity—an essential property for applications such as optimal pricing. }\label{tab1}
\end{table}

\begin{figure}[H]
\centering
\includegraphics[scale=0.5]{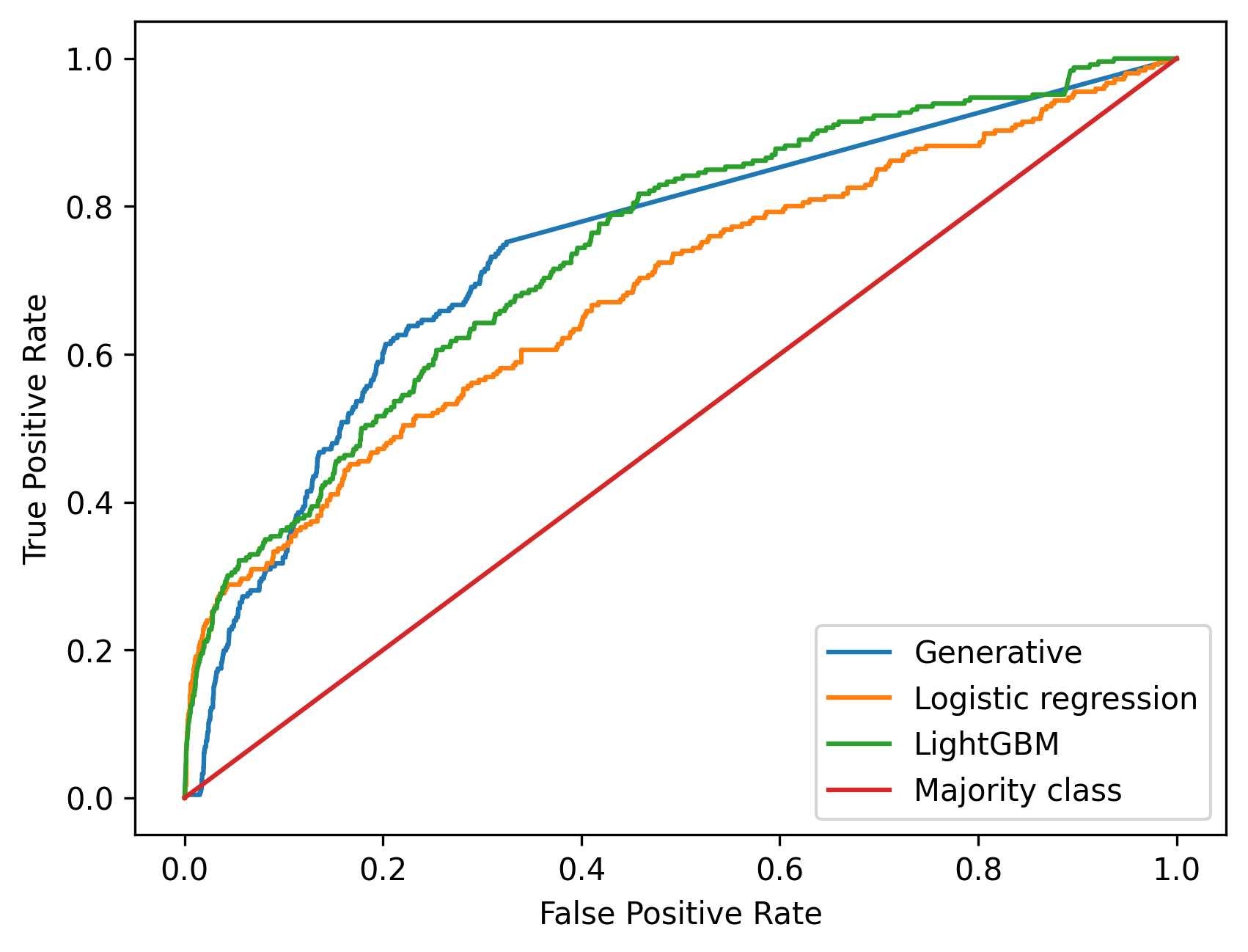}
\caption{ROC curves for the three evaluated models, including the majority class baseline. The generative model performs comparably to LightGBM, and both outperform logistic regression.}
\label{fig3}
\end{figure}

The comparative analysis of the three models reveals key insights into both predictive accuracy and the quality of probability estimates. LightGBM achieves the highest ROC-AUC at 0.743, but its advantage over the generative model (0.742) is marginal and not practically significant. Both models substantially outperform logistic regression, which attains a lower AUC of 0.684. These results are reinforced by inspecting the ROC curve in Fig~\ref{fig3}.

\begin{figure}[H]
    \centering
    \begin{subfigure}[b]{0.45\textwidth}
        \includegraphics[width=\textwidth]{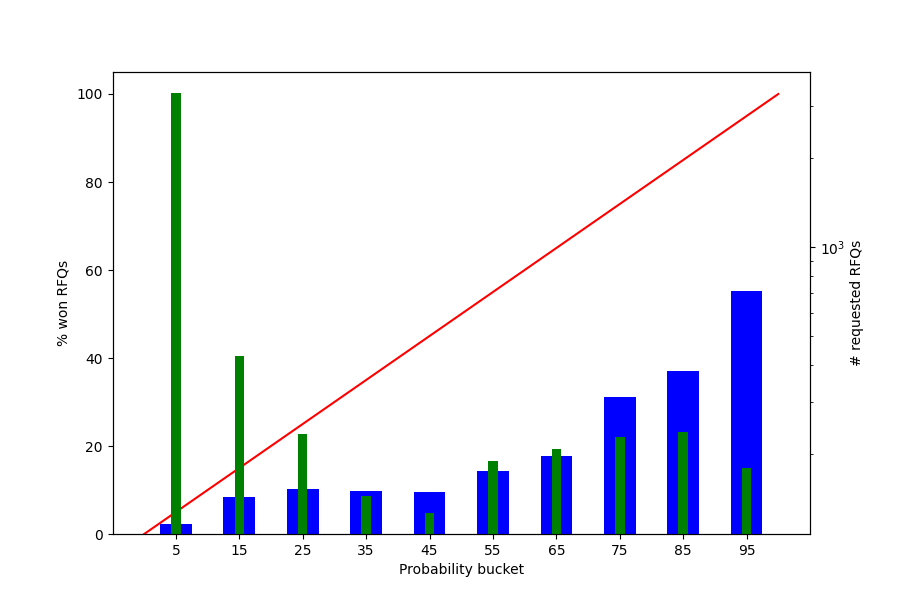}
        \caption{Generative}
        \label{fig4a}
    \end{subfigure}
    \hfill
    \begin{subfigure}[b]{0.45\textwidth}
        \includegraphics[width=\textwidth]{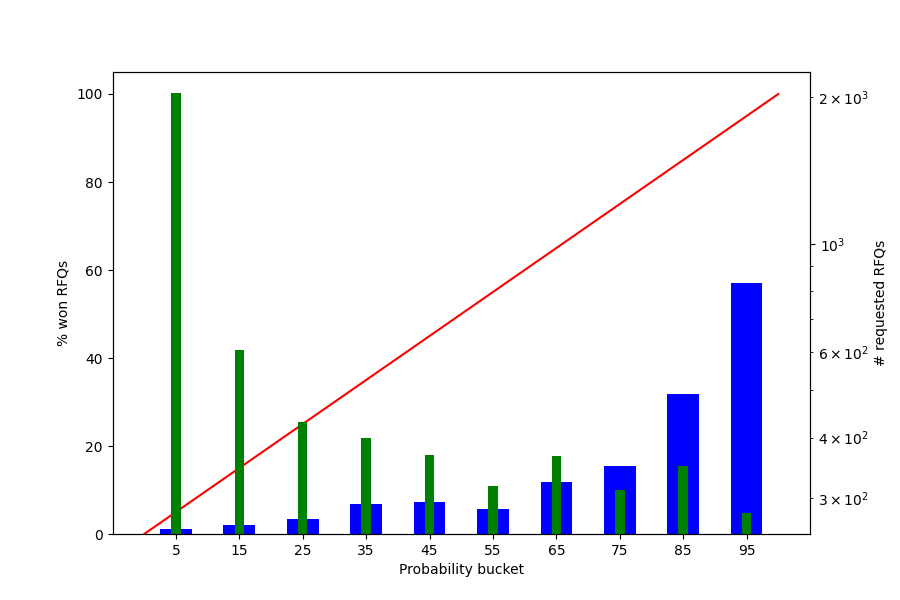}
        \caption{Logistic regression}
        \label{fig4b}
    \end{subfigure}

    \vspace{1em} 

    \begin{subfigure}[b]{0.45\textwidth}
        \includegraphics[width=\textwidth]{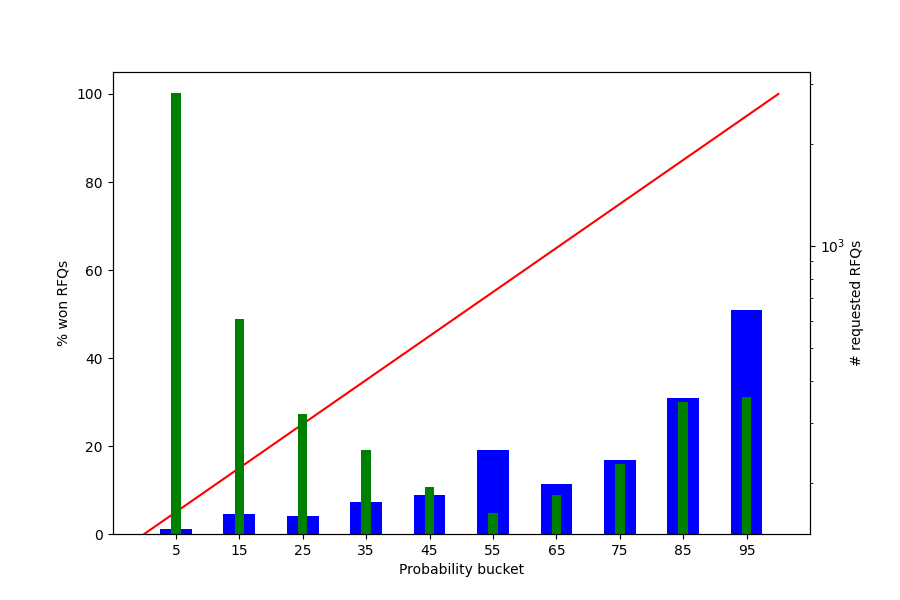}
        \caption{LightGBM}
        \label{fig4c}
    \end{subfigure}
    \hfill
    \begin{subfigure}[b]{0.45\textwidth}
        \includegraphics[width=\textwidth]{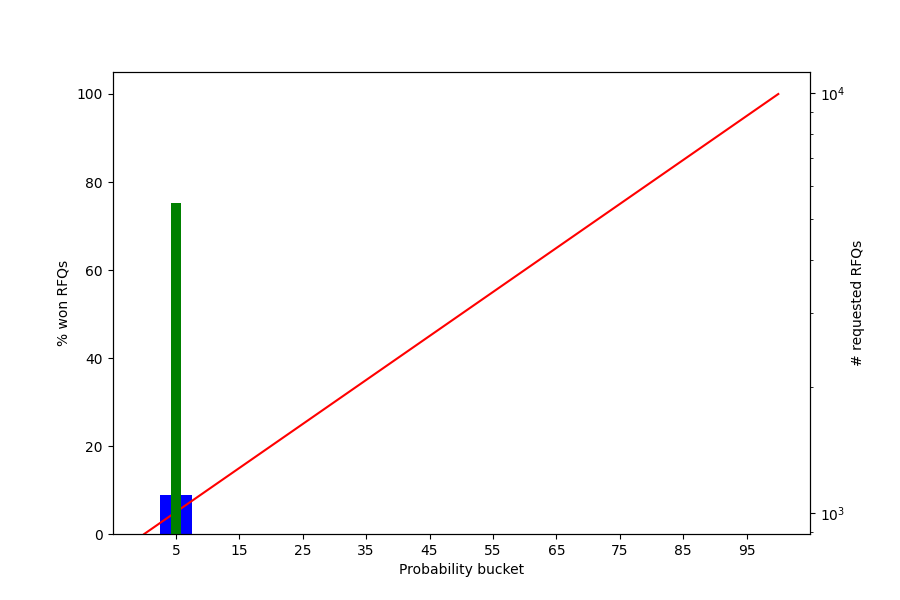}
        \caption{Majority class}
        \label{fig4d}
    \end{subfigure}

    \caption{Calibration plots for the three evaluated models and the majority class baseline. While the generative model appears smooth and close to the diagonal, class-balanced metrics indicate better calibration for logistic regression and LightGBM (see BBSS values in Table~\ref{tab1}).}
    \label{fig4}
\end{figure}

In terms of calibration, however, the generative model exhibits a lower Balanced Brier Skill Score (BBSS) than LightGBM and Logistic Regression, indicating less accurate probability estimates. This trade-off is partly attributable to the model’s distributional assumptions, which constrain its flexibility compared to discriminative methods. But also visual inspection of the calibration plots shown in Fig~\ref{fig4} suggests that the generative model produces smooth, nearly monotone probability estimates. However, once we evaluate calibration with class-balanced weighting, the ranking reverses. This indicates that the generative model’s calibration errors concentrate on the minority (hit) class: it looks well-calibrated on average, but once errors are weighted equally across classes, logistic regression and LightGBM provide more accurate probabilities.

\begin{figure}[H]
\centering
\includegraphics[scale=0.5]{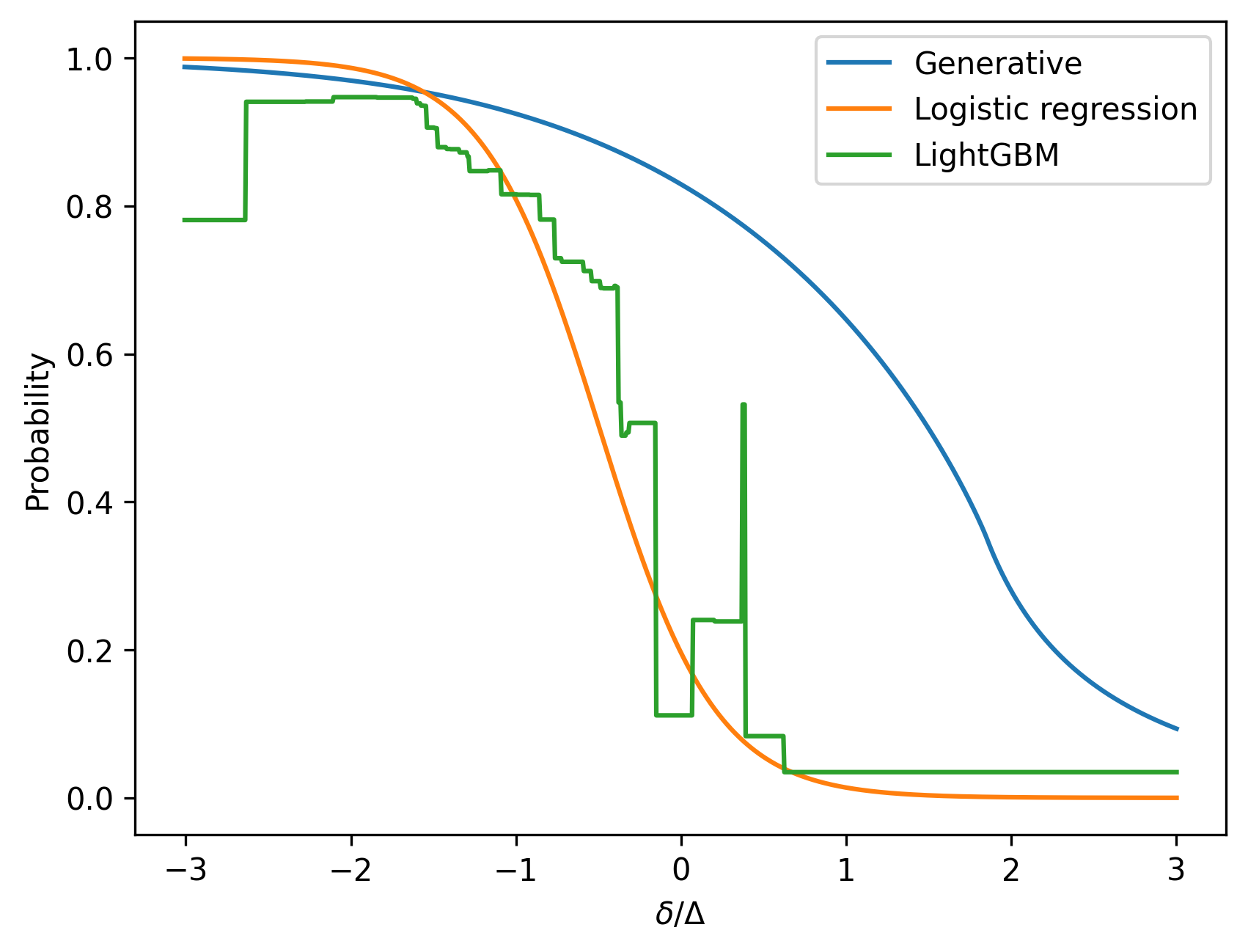}
\caption{Plot of predicted hit probability versus normalized half-spread for the three models, based on a specific RfQ. All input features are held constant except for the spread. As it can be clearly seen in the plot, the LightGBM model exhibits violations of monotonicity.
\label{fig5}}
\end{figure}

 Despite this, the generative framework provides an important practical advantage: it inherently enforces the business-critical constraint of spread monotonicity, ensuring that predicted hit probabilities increase as quoted spreads become more competitive. This property, shared by construction with the Logistic Regression model, is essential for economically consistent applications such as optimal pricing and dealer quote recommendations. As illustrated in Fig~\ref{fig5}, in contrast to the other two approaches, LightGBM exhibits erratic behavior in these areas, failing to reflect the expected increase in win probability as the quoted spreads become more attractive.

Overall, these findings underscore the benefits of modeling approaches that capture the intrinsic structure and logic of the underlying economic process. Although purely predictive models may achieve incremental gains through advanced architectures, hyperparameter optimization, or feature enrichment with post-trade information, the observed performance ceiling suggests that predictive accuracy alone is insufficient. In contrast, models that incorporate economic reasoning and structural constraints can deliver more robust and interpretable outcomes without sacrificing predictive power. This advantage is particularly salient in complex financial settings such as MD2C negotiation platforms—systems that are sufficiently intricate to pose meaningful modeling challenges, yet structured enough to permit explicit representation of their internal mechanisms.

In the case of the generative model, this strength arises from its explicit representation of the internal RfQ negotiation mechanics—capturing, for example, the simultaneous bidding behavior of multiple dealers and the client’s selection of the most competitive quote. Furthermore, its ability to integrate post-trade signals, such as the cover price or information about being the second-best bid, enhances both interpretability and performance, reinforcing the value of structurally grounded modeling strategies in financial applications.

One potential limitation of structurally informed models, however, lies in their inference time. Generative approaches, in particular, tend to be slower than the other models evaluated. This drawback may be mitigated by the fact that low-latency inference is typically more critical in highly liquid markets—precisely the contexts in which the relevance and applicability of structurally rich models are often more limited. This consideration, though, helps explain why simpler approaches, such as logistic regression, remain popular in practical applications \cite{JSG2024}.

To better understand the drivers of predictive performance, we examine the importance of variables across models. In Table \ref{tab2} we show the ranking of feature importance for each model. The majority class baseline does not use features, so it is not applicable. The ranking has been calculated using permutation importance for the generative model, standardized coefficients for the LR, and gain-based feature importance for LightGBM. More details are available in \nameref{S3_Table}.

\begin{table}[H]
\centering
\begin{tabular}{|l|c|c|c|}
\hline
\textbf{Variable} & \textbf{Generative} & \textbf{Logistic R} & \textbf{LightGBM} \\ \hline
Spread            & 1 & 1 & 1 \\ \hline
Volatility        & 3 & 7 & 8 \\ \hline
DV01              & 5 & 2 & 7 \\ \hline
Frequency buy     & 7 & 8 & 6 \\ \hline
Frequency sell    & 6 & 6 & 9 \\ \hline
Average dealers   & 4 & 4 & 5 \\ \hline
Maturity          & 2 & 9 & 4 \\ \hline
Number of dealers & 8 & 5 & 3 \\ \hline
DV01 exposure     & 9 & 3 & 2 \\ \hline
\end{tabular}
\caption{Ranking of feature importance across models for sell RfQs}
\label{tab2}
\end{table}

Across models, spread is the dominant driver of hit probability: it ranks first in LightGBM, the logistic model, and the generative model. The second tier is led by risk/size proxies: DV01 (logistic) and DV01 exposure (LightGBM) carry substantial importance, indicating that larger risk‐adjusted notionals materially affect execution. Microstructure controls—number of dealers, maturity, volatility, and client frequency—play a secondary role: LightGBM assigns non-trivial importance to number of dealers (competition intensity) and to maturity (liquidity/duration risk), while the generative specification picks up maturity and volatility more than the frequency variables. The overall pattern is consistent: pricing aggressiveness (spread) and risk/size (DV01) are the primary drivers of execution, with platform and liquidity frictions shaping outcomes at the margin.

\section{Conclusions}

This work has presented a unified framework for understanding and optimizing dealer behavior in electronic bond trading on Multi-Dealer-to-Client (MD2C) platforms, grounded in the formal tools of probabilistic graphical models and causal inference. By representing the RfQ process through a causal graph, we uncover latent structures and confounding relationships that often undermine traditional data-driven approaches to business interventions such as pricing, client targeting, and profitability estimation. Our framework supports a rigorous analysis of these interventions, enabling the identification of conditions under which valid causal reasoning can be applied—leveraging the rich historical datasets typically available to dealers. In the absence of such tools, dealers risk deploying suboptimal strategies that fail to account for underlying causal dynamics. A summary of a typical workflow using our framework is provided in Fig~\ref{fig6}. 

\begin{figure}[H]
\centering
\begin{tikzpicture}[
  font=\small,
  node distance=6mm,
  box/.style={draw, rounded corners, align=center, inner sep=3pt, fill=gray!5, text width=80mm},
  line/.style={-{Latex[length=2mm]}, thick}
]
\node[box] (inputs) {\textbf{Observational Data }\\ RfQ (side $s$, dealers $n$, half-spread $\delta$, volume $v$) \\ Market state (drift $\mu$, volatility $\sigma$, BF, CF)};
\node[box, below=of inputs] (causal) {\textbf{Causal Inference}\\ Graph \& do-calculus; backdoor; select conditioning set $\mathcal{Z}_t$};
\node[box, below=of causal] (models) {\textbf{Model Specification}\\ Generative; Discriminative (Logistic Regression, LightGBM)};
\node[box, below=of models] (estim) {\textbf{Model Estimation}\\ Maximum Likelihood};
\node[box, below=of estim] (eval) {\textbf{Evaluation \& Model Selection}\\ ROC/AUC, BBSS, Business Constraints};
\node[box, below=of eval] (interv) {\textbf{Optimal pricing policy}\\ $do(\delta^*)$};
\node[below=0.5cm of interv] {}; 
\draw[line] (inputs) -- (causal);
\draw[line] (causal) -- (models);
\draw[line] (models) -- (estim);
\draw[line] (estim) -- (eval);
\draw[line] (eval) -- (interv);
\end{tikzpicture}
\caption{Overview of the frameworks developed in this work for causal interventions on MD2C platforms, exemplified through the optimal pricing problem. We apply causal inference to identify the conditioning set $\mathcal{Z}_t$ that enables reduction of the problem to empirically estimable probabilities—specifically, the hit probability model. Both generative and discriminative model formulations are used to estimate this probability, with parameters fitted via maximum likelihood. Their performance is then evaluated on a separate validation set using standard predictive metrics tailored to this problem, while incorporating business constraints such as monotonicity with respect to the spread. Finally, the selected model is employed to compute the optimal spread, denoted $\delta^*$}
\label{fig6}
\end{figure}

Building upon this framework, we have analyzed several key business applications: optimal pricing strategies under various revenue definitions, a classification model for assessing RfQ revenue potential, and an uplift-based axe matcher to identify clients likely to respond to commercial actions. Crucially, applying the framework to these use cases reveals that neglecting confounders—such as bond and client characteristics—or failing to account for partially observable factors, including information asymmetry or client intent (e.g., price discovery), can bias predictive models and ultimately result in suboptimal decision-making.

We implemented both generative and discriminative modeling strategies to estimate the causal effects underlying these interventions, aiming to assess their respective strengths in extracting actionable insights from the proposed framework. The generative model—built upon and extending the foundational work of Fermanian et al.—captures key structural aspects of the RfQ process, including competitive dynamics, client reservation spreads, and post-trade signals such as the cover price. It adheres to fundamental economic assumptions, such as monotonicity with respect to spread, and enables fully probabilistic reasoning. In contrast, the discriminative models—logistic regression and LightGBM—are designed to optimize predictive accuracy but do not inherently encode domain-specific constraints, except in cases like logistic regression, where such constraints may be satisfied by construction.

An empirical evaluation based on a proprietary dataset of BBVA’s Request-for-Quote (RfQ) activity in European government bonds shows that the generative model achieves performance comparable to LightGBM in terms of AUC. The logistic regression model, while offering straightforward interpretability and satisfying monotonicity by construction, underperforms on classification metrics. This contrast is particularly noteworthy given that LightGBM—a state-of-the-art machine learning approach—does not impose monotonicity constraints by default. The trade-off for the generative model’s structural rigor is a deterioration in calibration metrics, as its reduced flexibility limits its ability to produce well-calibrated probabilities across the entire range. Nevertheless, for the dealer intervention use cases discussed in this study, we consider this trade-off worthwhile, since violating structural constraints such as monotonicity can lead to economically inconsistent or unreliable decisions in practice.

These results suggest that embedding structural knowledge of the RfQ process directly into the model provides advantages that are not easily replicated through predictive optimization alone. While further enhancements to discriminative models—such as advanced feature engineering or the use of constrained architectures—remain possible, our findings highlight the value of hybrid or generative approaches that incorporate domain-specific structure and are grounded in causal reasoning. This has implications beyond this particular setup, offering a concrete example of a complex economic environment where the trade-offs between purely data-driven and model-driven approaches can be meaningfully evaluated.

Several promising directions emerge for future work, offering pathways to extend and validate the proposed framework. One line of research focuses on enhancing the predictive performance of discriminative models by making full use of the available information, while also adhering to key business constraints. This includes developing models that incorporate domain-specific requirements—such as monotonic neural networks—and enhancing feature engineering to integrate informative post-trade signals like cover prices. This work is ongoing; see \cite{paloma_thesis} for further developments. Other recent studies have explored alternative feature engineering approaches, including quantum-enhanced representations that aim to enrich probabilistic models with higher-order informational structure \cite{ciceri2025enhanced}.

Given the proprietary nature of the dataset used in this study, an important direction for future work is cross-institutional validation. Applying the proposed framework to RfQ data from other institutions—even if non-public—would help evaluate its generality and robustness across different dealer environments and market structures. Our results indicate that the causal intervention framework enhances predictive modeling in bond RfQ platforms. When the trading mechanism remains the same—i.e., an RfQ-based dealer-to-client process—the framework should be overall portable; what requires adaptation is the conditioning set and, potentially, the parameterization. Extending the approach to other asset classes, such as corporate bonds or derivatives, will therefore necessitate re-tuning the instrument characteristics, and client profiles. For example, the conditioning variables may need to incorporate credit ratings, alternative risk measures (e.g., CDS spreads, duration/DV01), collateral eligibility, margining practices, or contract-specific features (e.g., option Greeks). In some cases, minor variations in the RfQ trading protocol may need to be adapted to the specific use case. In summary, while the graphical model is, in principle, flexible enough to accommodate most of these elements, a careful reassessment would still be required before applying the framework to new contexts.

While our empirical evaluation focused primarily on optimal pricing, additional causal interventions explored in this work—such as revenue potential estimation and uplift-based axe matching—warrant further empirical analysis. However, progress in this area is often constrained by limited availability of detailed records on commercial actions, such as client outreach and follow-ups, even within proprietary datasets.

A particularly valuable extension involves incorporating the proposed causal framework into the multi-RfQ optimization problem, where dealers must determine optimal pricing for a given RfQ while anticipating the arrival of other RfQs in the same direction or with different volumes, prior to the final liquidation of the position. This setting reflects the real-world complexity of managing partially executed positions and requires optimizing not only for the current trade, but also for its interaction with future flow. Addressing this problem effectively calls for an integration of the causal inference tools developed in this work with stochastic optimal control techniques, such as those introduced by Avellaneda and Stoikov. While their framework already embeds a form of causal reasoning, it does not fully capture the structural richness and latent complexity addressed in the present work. Combining these perspectives could yield a more robust and realistic foundation for dynamic decision-making in multi-RfQ environments.

Finally, the graphical model introduced here offers a flexible foundation for reasoning about a wider range of probabilistic and causal questions beyond the specific applications discussed. As such, it provides a principled and extensible basis for rigorous, data-driven decision-making in electronic bond trading via Multi-Dealer-to-Client platforms.

\nolinenumbers

%
%

\bibliography{bibliografia}

\section*{Supporting information}

\paragraph*{S1 Appendix.}
\label{S1_Appendix}
\textbf{Data preprocessing.} Complete description of cleaning, filters, and outlier removal.

\paragraph*{S1 Table.}
\label{S1_Table}
\textbf{Data summary table.} Summary statistics of the RFQ dataset (size, period, sides, dealers, notional, frequency).

\paragraph*{S2 Table.}
\label{S2_Table}
\textbf{Notation.} Reference table.

\paragraph*{S3 Table.}
\label{S3_Table}
\textbf{Feature importance.} Feature importance across models for sell RfQs.

\end{document}